\documentclass[%
  conference,%
  10pt,%
]{IEEEtran}

\usepackage{cite}
\usepackage{amsmath,amssymb,amsfonts}
\allowdisplaybreaks%

\usepackage{booktabs}
\usepackage{url}

\usepackage{paralist}

\usepackage[UKenglish]{babel}

\usepackage{graphicx}
\graphicspath{{./images/}}
\usepackage{placeins}
\usepackage{subcaption}
\usepackage{enumitem}
\usepackage{balance}
\usepackage{booktabs}  
\usepackage{colortbl}

\usepackage[%
  group-minimum-digits=4,%
  list-final-separator={, and },%
  add-integer-zero=false,%
  free-standing-units,%
  binary-units,%
  detect-weight=true,%
  detect-inline-weight=math,%
]{siunitx}
\usepackage{microtype}

\usepackage[colorinlistoftodos,prependcaption,textsize=tiny]{todonotes}

\usepackage{xspace}

\usepackage[capitalise]{cleveref}

\crefformat{footnote}{#2\footnotemark[#1]#3}

\usepackage{tabularx}

\newcommand{\summary}[2]{
    \vspace{1em}
    \noindent
    \colorbox{gray!20}{%
        \parbox{.97\linewidth}{%
                \textbf{Summary \textit{(#1)}}
            #2
        }%
    }%
}%

\usepackage{csquotes}
\usepackage{csvsimple}
\usepackage{listings}
\usepackage{xcolor}

\definecolor{codegreen}{rgb}{0,0.6,0}
\definecolor{codegray}{rgb}{0.5,0.5,0.5}
\definecolor{codepurple}{rgb}{0.58,0,0.82}
\definecolor{backcolour}{rgb}{0.95,0.95,0.92}

\lstdefinestyle{mystyle}{
  backgroundcolor=\color{backcolour},
  commentstyle=\color{codegreen},
  keywordstyle=\color{magenta},
  numberstyle=\tiny\color{codegray},
  stringstyle=\color{codepurple},
  basicstyle=\ttfamily\footnotesize,
  breakatwhitespace=false,
  breaklines=true,
  breakindent=10pt,
  captionpos=b,
  keepspaces=true,
  numbers=left,
  numbersep=5pt,
  showspaces=false,
  showstringspaces=false,
  showtabs=false,
  tabsize=2
}

\newcommand\copyrighttext{%
  \footnotesize
Accepted at ICST2022.\@ \copyright~2022 IEEE.
Personal use of this material is permitted.
Permission from IEEE must be obtained for all other uses, in any current or future media, including
reprinting/republishing this material for advertising or promotional purposes, creating new collective works,
for resale or redistribution to servers or lists, or reuse of any copyrighted component of this work in other works.
}
\newcommand\copyrightnotice{%
    \begin{tikzpicture}[remember picture,overlay]
        \node[anchor=south,yshift=10pt] at (current page.south) {\fbox{\parbox{\dimexpr\textwidth-\fboxsep-\fboxrule\relax}{\copyrighttext}}};
    \end{tikzpicture}%
}

\begin{document}

\title{A Survey on How Test Flakiness Affects Developers and What Support They Need To Address It}

\author{%
 \IEEEauthorblockN{Martin Gruber}%
 \IEEEauthorblockA{%
   \textit{BMW Group, University of Passau}\\%
   Munich, Germany\\%
   martin.gr.gruber@bmw.de%
 }%
 \and%
 \IEEEauthorblockN{Gordon Fraser}%
 \IEEEauthorblockA{%
   \textit{University of Passau}\\%
   Passau, Germany\\%
   gordon.fraser@uni-passau.de%
 }
}

\maketitle

\newcommand{\figshareURL}{\url{https://doi.org/10.6084/m9.figshare.16727251}}
\newcommand{\myplus}{\oplus}
\newcommand{\myminus}{\ominus}

\newcommand{\prevalenceRQ}{RQ1\xspace}
\newcommand{\rootCauseRQ}{RQ2\xspace}
\newcommand{\consequencesRQ}{RQ3\xspace}
\newcommand{\mitigationsRQ}{RQ4\xspace}
\newcommand{\wishesRQ}{RQ5\xspace}

\newcommand{\questionPlatform}{For which platforms do you develop?}
\newcommand{\questionSoftwareType}{What types of software do you develop?}
\newcommand{\questionCodeSize}{How large is the code base of your main project by lines of source code (LOC)?}
\newcommand{\questionTeamSize}{How many developers are working on your product or project?}
\newcommand{\questionTestLevel}{What kinds of tests do you typically work with?}
\newcommand{\questionTestingPractices}{Which of these testing practices do you use?}
\newcommand{\questionFrequency}{How often do you experience flaky failures?}
\newcommand{\questionSeverity}{How problematic is flakiness for you?}
\newcommand{\questionConsequences}{Which negative effects of flaky tests have you experienced?}
\newcommand{\questionRootCause}{What are the root causes of the flakiness you experienced?}
\newcommand{\questionMitigations}{What is your current strategy to deal with flakiness?}
\newcommand{\questionWishes}{What tool or information would you like to have in order to help you deal with flaky tests?}

\newcommand{\numBothFreqNever}{13\xspace}

\newcommand{\percentAutomotiveConsequenceMergingRRsVeryoftenAlways}%
    {\SI{76}{\percent}\xspace}

\newcommand{\percentAutomotiveRootCauseIONever}{\SI{71}{\percent}\xspace}

\newcommand{\preSurveyNumOfSubmissionsBeforePreproc}{612\xspace}
\newcommand{\preSurveyNumOfSubmissionsWithoutSameAnswerMultiSubmissions}{607\xspace}
\newcommand{\preSurveyNumOfSubmissionsUnique}{604\xspace}
\newcommand{\preSurveyNumOfSubmissionsTrippleYes}{302\xspace}

\newcommand{\globalNumFullyCompleteSubmissions}{228\xspace}
\newcommand{\globalNumNotFullyIncompleteSubmissions}{233\xspace}

\newcommand{\automotiveNumFullyCompleteSubmissions}{73\xspace}
\newcommand{\automotiveNumNotFullyIncompleteSubmissions}{102\xspace}

\newcommand{\bothNumFullyCompleteSubmissions}{301\xspace}
\newcommand{\bothNumNotFullyIncompleteSubmissions}{335\xspace}

\newcommand{\globalAgeMin}{18\xspace}
\newcommand{\globalAgeMax}{63\xspace}
\newcommand{\globalAgeMean}{29.5\xspace}
\newcommand{\globalAgeMedian}{28\xspace}

\newcommand{\globalNationalityPercentRomania}{\SI{0}{\percent}\xspace}
\newcommand{\globalNationalityPercentPakistan}{\SI{0}{\percent}\xspace}
\newcommand{\globalNationalityPercentFinland}{\SI{0}{\percent}\xspace}
\newcommand{\globalNationalityPercentBelgium}{\SI{0}{\percent}\xspace}
\newcommand{\globalNationalityPercentMorocco}{\SI{0}{\percent}\xspace}
\newcommand{\globalNationalityPercentVenezuelaBolivarianRepublicof}{\SI{0}{\percent}\xspace}
\newcommand{\globalNationalityPercentChina}{\SI{0}{\percent}\xspace}
\newcommand{\globalNationalityPercentSlovenia}{\SI{0}{\percent}\xspace}
\newcommand{\globalNationalityPercentArgentina}{\SI{0}{\percent}\xspace}
\newcommand{\globalNationalityPercentZimbabwe}{\SI{0}{\percent}\xspace}
\newcommand{\globalNationalityPercentLatvia}{\SI{0}{\percent}\xspace}
\newcommand{\globalNationalityPercentSweden}{\SI{0}{\percent}\xspace}
\newcommand{\globalNationalityPercentCostaRica}{\SI{0}{\percent}\xspace}
\newcommand{\globalNationalityPercentNorway}{\SI{0}{\percent}\xspace}
\newcommand{\globalNationalityPercentBrazil}{\SI{1}{\percent}\xspace}
\newcommand{\globalNationalityPercentFrance}{\SI{1}{\percent}\xspace}
\newcommand{\globalNationalityPercentNepal}{\SI{1}{\percent}\xspace}
\newcommand{\globalNationalityPercentCzechRepublic}{\SI{1}{\percent}\xspace}
\newcommand{\globalNationalityPercentChile}{\SI{1}{\percent}\xspace}
\newcommand{\globalNationalityPercentTurkey}{\SI{1}{\percent}\xspace}
\newcommand{\globalNationalityPercentIsrael}{\SI{1}{\percent}\xspace}
\newcommand{\globalNationalityPercentEstonia}{\SI{1}{\percent}\xspace}
\newcommand{\globalNationalityPercentGermany}{\SI{2}{\percent}\xspace}
\newcommand{\globalNationalityPercentIndia}{\SI{2}{\percent}\xspace}
\newcommand{\globalNationalityPercentHungary}{\SI{2}{\percent}\xspace}
\newcommand{\globalNationalityPercentNetherlands}{\SI{2}{\percent}\xspace}
\newcommand{\globalNationalityPercentCanada}{\SI{3}{\percent}\xspace}
\newcommand{\globalNationalityPercentGreece}{\SI{3}{\percent}\xspace}
\newcommand{\globalNationalityPercentMexico}{\SI{3}{\percent}\xspace}
\newcommand{\globalNationalityPercentSpain}{\SI{5}{\percent}\xspace}
\newcommand{\globalNationalityPercentUnitedKingdom}{\SI{5}{\percent}\xspace}
\newcommand{\globalNationalityPercentSouthAfrica}{\SI{6}{\percent}\xspace}
\newcommand{\globalNationalityPercentItaly}{\SI{7}{\percent}\xspace}
\newcommand{\globalNationalityPercentUnitedStates}{\SI{10}{\percent}\xspace}
\newcommand{\globalNationalityPercentPoland}{\SI{11}{\percent}\xspace}
\newcommand{\globalNationalityPercentPortugal}{\SI{25}{\percent}\xspace}
\newcommand{\globalNationalityPercentSouthAmerica}{\SI{3}{\percent}\xspace}
\newcommand{\globalNationalityPercentAsia}{\SI{6}{\percent}\xspace}
\newcommand{\globalNationalityPercentAfrica}{\SI{6}{\percent}\xspace}
\newcommand{\globalNationalityPercentNorthAmerica}{\SI{17}{\percent}\xspace}
\newcommand{\globalNationalityPercentEurope}{\SI{69}{\percent}\xspace}

\newcommand{\globalSexPercentageMale}{\SI{81.1}{\percent}\xspace}
\newcommand{\globalSexPercentageFemale}{\SI{18.9}{\percent}\xspace}

\newcommand{\globalStudentStatusYes}{\SI{35}{\percent}\xspace}
\newcommand{\globalStudentStatusNo}{\SI{64}{\percent}\xspace}

\newcommand{\globalEmploymentStatusFullTime}{\SI{80.3}{\percent}\xspace}
\newcommand{\globalEmploymentStatusPartTime}{\SI{19.7}{\percent}\xspace}

\begin{abstract}

Non-deterministically passing and failing test cases, so-called \textit{flaky}
tests, have recently become a focus area of software engineering research.
While this research focus has been met with some enthusiastic endorsement from
industry, prior work nevertheless mostly studied flakiness using a code-centric
approach by mining software repositories.
What data extracted from software repositories cannot tell us, however, is how
developers perceive flakiness: How prevalent is test flakiness in developers'
daily routine, how does it affect them, and most importantly: What do they want
us researchers to do about it?
To answer these questions, we surveyed \bothNumNotFullyIncompleteSubmissions
professional software developers and testers in different domains.
The survey respondents confirm that flaky tests are a common and serious
problem, thus reinforcing ongoing research on flaky test detection. Developers are less worried about the computational costs caused by re-running tests and more about the loss of trust in the test outcomes.
Therefore, they would like to have IDE plugins to detect flaky code
as well as better visualizations of the problem, particularly dashboards showing
test outcomes over time; they also wish for more training and information on
flakiness. These important aspects will require the attention of researchers as well as tool developers.

\end{abstract}

\copyrightnotice
\begin{IEEEkeywords}
  Flaky Tests; Empirical Study; Survey
\end{IEEEkeywords}

\section{Introduction}%
\label{sec:introduction}

Flaky tests are tests that behave non-deterministically, so they may both pass and fail when executed on the same code under test repeatedly.
The problem of flaky tests has been extensively researched in the recent past, but prior work~\cite{Parry2021} primarily used a code-centric approach, providing only limited insights into how test flakiness affects developers.
Furthermore, the few human studies that do exist never asked their participants directly about their wishes towards researchers and tool developers.

In order to address these limitations, we aim to complement prior research by finding out from developers in different fields directly how they experience flakiness, what they
have already tried to mitigate it, and what tools or information they
would desire to better handle this challenge.
We surveyed \bothNumNotFullyIncompleteSubmissions professional
 developers on their experiences with flaky tests and what they
 desire from the research community. Our sample population consists of
\globalNumNotFullyIncompleteSubmissions professionals from the general public all over the globe, as
well as \automotiveNumNotFullyIncompleteSubmissions employees of the BMW group. Using this setup, we are able to investigate flakiness in
both a specific domain and software development in general.
In detail, by asking this large set of developers, we aim to shed light on the
following questions:

\noindent\textbf{Prevalence and Severity:}
We asked participants how they experience flakiness.
Developers perceive flakiness as a common and severe issue, with \SI{51}{\percent} of all participants experiencing it at least weekly and \SI{66}{\percent} rating it as a moderate or serious problem.
Despite ranking flakiness a less severe issue, mobile application developers experience flaky failures very often due to flaky emulators and testing tools.
On the other hand, safety-relevant software running in cars sees flaky failures less often but understandably considers it a more severe problem.
Overall, flakiness is perceived as an equally or more pressing challenge in the automotive domain than in other domains.

\noindent\textbf{Root Causes:}
We asked participants what causes of flakiness they have experienced.
While concurrency-related issues remain overall most common, we observed differences between domains and also discovered uninitialized variables, compiler differences, and undocumented API changes to cause test flakiness---three
previously undocumented categories.

\noindent\textbf{Consequences:}
We asked participants how flakiness affects them. Losing trust and
wasting developer time are perceived as the most severe impact,
particularly more than the waste of computational resources through test reruns.
Furthermore, we showed that the trust developers put in a test outcome can also depend on the outcome itself, which is important for tool developers who are already building on this assumption~\cite{Machalica2020}.

\noindent\textbf{Mitigations:}
When asked how they currently handle flaky tests, developers responded that
they mostly fall back to rerunning and rewriting test cases with little use of
automated techniques; this is in part because there is a lack of practical tools for many development environments.

\noindent\textbf{Wishes:}
Finally, we asked developers directly what tools or information they would like to possess to better deal with flakiness.
They wish to use more sophisticated technologies such as IDE plugins that can statically detect flaky tests, as well as better visualizations of flakiness, like showing the outcomes of a test case over time.
Lastly, they want more educational opportunities such as examples, guides, and best practices on how to deal with flaky tests.
This knowledge should help tool developers to achieve a higher impact of their work.

It is our hope that these insights help to build a clear vision of how future research on test flakiness can create value for developers.
We provide all data freely, allowing other researchers to verify, replicate, and extend our findings:\newline\figshareURL

\section{background}%
\label{sec:background}

\subsection{Prevalence and Severity}%
\label{sec:prevalence_and_severity}

Scientific studies have found \SI{0.5}{\percent}~\cite{Lam2019} to \SI{1}{\percent}~\cite{Gruber2021} of tests to be flaky.
Practitioners at Google report a similar diffusion of flakiness among small tests~\cite{Listfield2017} and a much stronger prevalence in larger ones:
Overall, almost \SI{16}{\percent} of their tests show some kind of flakiness, and \SI{84}{\percent} of the transitions from pass to fail involve a flaky test~\cite{Micco2016}; out of \num{115160} test targets that had previously passed and failed at least once, \SI{41}{\percent} were flaky~\cite{Memon2017}.
These technical metrics, however, do not necessarily reflect how often developers have to deal with the problem.

At the time of our study, there was only one other survey asking developers directly about their experience with flakiness, which involved 121 developers and was conducted by Eck et al.~\cite{Eck2019} in 2019.
Of their participants, \SI{58}{\percent} indicated that they have to deal with flaky tests at least monthly.
Out of the \SI{91}{\percent} experiencing the issue at least a few times a year, \SI{79}{\percent} rated it a moderate or serious problem.
Eck et al.\ pointed out the importance of replications to corroborate their findings and the need for further research assessing the role played by additional factors like project and organizational domains.
We are not only following this call but go beyond existing experience reports in both number of participants and detail of questioning.

\subsection{Root Causes}%
\label{sec:background_root_causes}
Luo et al.~\cite{Luo2014} first investigated flaky tests by inspecting 201 commits that likely fix flaky tests in 51 open-source Apache projects.
They identified ten possible root causes for flakiness, of which concurrency, especially asynchronous waiting, was the most prominent one.
Later studies confirmed and extended this list by inspecting test-related bug reports~\cite{Vahabzadeh2015}, commits relevant to flakiness in Android projects~\cite{Thorve2018}, previously fixed flaky tests extracted from the Mozilla bug tracker~\cite{Eck2019}, or pull requests fixing flaky tests in Microsoft projects~\cite{Lam2020}.
The main limitation of these studies is that they identify and quantify root causes only by investigating fixes of flakiness, but the number of issues or pull requests concerning flaky tests might not correspond to how often developers have to deal with a specific type of flakiness.
This concern is supported by the fact that the one study taking a different approach by identifying flaky tests through reruns and classifying them by inspecting their code~\cite{Gruber2021} disagrees with previous studies finding networking and randomness to be the most prevalent causes.
With our survey, we search for the causes of flakiness that are most prominent from the developers' point of view.

\subsection{Consequences}%
\label{sec:background_consequences}

Trying to quantify the impact of flaky tests, Rahman and Rigby~\cite{Rahman2018} showed that Firefox builds containing failing flaky tests receive more crash reports than average builds.
Other studies showed that even if flaky tests are rare, they can still cause many build failures~\cite{Lam2019a, Lam2020, Labuschagne2017} and that up to \SI{16}{\percent} of computational resources are spent rerunning flaky tests~\cite{Micco2017}.
Such metrics can serve as proxies to measure the impact of test flakiness, although its most commonly stated effects, namely \textit{losing trust in testing} and \textit{wasting developer time}, have only rarely been studied~\cite{Eck2019}.
In our survey, we compare different forms of resource waste, trust loss, and other impediments.

The loss of trust in tests has been reported to be
asymmetric~\cite{Machalica2020}: Passing tests are seen as an indication for
the absence of regression, whereas failures are seen as a hint to run the test
again. This \textit{idiosyncrasy of software development}~\cite{Machalica2020}
is an assumption often underlying flakiness-detection techniques. While hints
for this phenomenon can often be found in the literature (e.g.,
\enquote*{You'll need to run a test suite multiple times to determine whether
you have a bug or a flaky test}~\cite{Oezal2021a}), these are only individual
examples that indicate the presence of such an asymmetric perception. Using
statistical tests, we aim to scientifically validate the existence of this
phenomenon.

\subsection{Mitigation Strategies}%
\label{sec:background_mitigations}

Similar to bugs, flakiness is considered a problem that cannot be eradicated completely~\cite{Raine2020}. For example, despite Google's efforts to reduce test flakiness, they are facing a constant rate of \SI{1.5}{\percent} of all test executions producing flaky results~\cite{Micco2016}.
To mitigate the problem, multiple approaches have been proposed:
\begin{inparaenum}
	\item \textbf{Rerun \& Disable:} This is the simplest way to deal with flakiness, but it wastes resources and leads to delay or loss of test feedback.
	\item \textbf{Flag:} Automatically rerun tests via annotations~\cite{Wendelin, PytestFlakyRerunPlugin, JenkinsFlakyTestHandlerPlugins}.
	\item \textbf{Auto Detect:} Minimize the number of reruns by inspecting version control histories (DeFlaker)~\cite{Bell2018}, smart test shuffling (iDFlakies)~\cite{Lam2019}, identifying test dependencies (PraDet)~\cite{Gambi2018}, or predicting test flakiness via static~\cite{Bertolino2020, Pinto2020}, or hybrid~\cite{Alshammari2021} analysis.
	\item \textbf{Auto Debug:} Find root causes by comparing runtime parameters~\cite{Lam2019a} or coverage behavior~\cite{Ziftci2020}, and propose fixes for order-dependent tests~\cite{Shi2019} or asynchronous waiting~\cite{Lam2020}.
	\item \textbf{Incentives \& Penalties:} Strong management oversight, assign team members and time slots to fix flaky tests~\cite{Champier2019}.
	\item \textbf{Quantify \& Visualize:} Create attention for the problem~\cite{Wendelin}, outline expected reward for fixing~\cite{Machalica2020}, and use dashboards to visualize flakiness~\cite{Oezal2021a, Champier2019, Liviu2019, Palmer2019}.
	\item \textbf{Auto report \& disable:} Submit bug reports and move tests to separate test suite (\textit{Quarantining})~\cite{Fowler2011, Wendelin, Micco2016, Lam2020}.
\end{inparaenum}
To the best of our knowledge, there is no study on how frequently these
techniques are being applied. We aim to close this gap in order to provide
researchers valuable feedback on the adoption rates of their approaches.

\subsection{Wishes}%
\label{sec:background_wishes}

So far, little is known about the developers' desires regarding what imaginable defense mechanisms against test flakiness they would like to possess.
In their survey, Eck et al.~\cite{Eck2019} provided participants with a list of eight literature-selected pieces of information and asked which of those are most important and most difficult to obtain in order to fix a flaky test and what additional challenges they face.
They found that reproducing the context leading to the test failure and understanding the nature of the flakiness are the most important and challenging needs and that designing test code properly to avoid flakiness is an additional challenge.
Unlike them, we ask developers a more open question, hoping to collect creative ideas that are not restricted by existing literature.

\section{Study Setup}%
\label{sec:study_setup}

Through our study, we aim to empirically answer the following research questions:
\begin{description}

    \item[\prevalenceRQ:] How prevalent and severe is the problem of flakiness?

    \item[\rootCauseRQ:] What are the causes of flakiness?

    \item[\consequencesRQ:] What are the consequences of flakiness?

    \item[\mitigationsRQ:] What are common strategies to mitigate flakiness?

    \item[\wishesRQ:] What tools or information would developers wish for to better handle flaky tests?

\end{description}

\subsection{Target Audience and Participant Selection}%
\label{sec:target_audience_and_participant_selection}

Our target audience are professional software developers and testers since they experience potentially existing flakiness in its full extent and unfiltered form.
Picking a sampling strategy for this population is challenging since one has to balance two trade-offs:
Limiting the population to developers working for only one (or a few) companies paints a detailed picture and controls for company-specific variables like policies or organizational structures, which all participants have in common.
However, the resulting findings might not generalize beyond that company.
On the other hand, broadening the scope by sampling participants from different companies, countries, and industries creates a more general overview, although the results might be influenced by unknown variables.
By conducting our study on two different sample populations, one from a specific company and one global, we hope to address these concerns and therefore strengthen both internal and external validity of our survey.
In the following, we refer to these two groups as \textit{automotive} and \textit{global} participants.

\subsubsection{Automotive Survey}%
\label{sec:automotive_survey}

All automotive participants work with the BMW group.
To promote our survey, we contacted teams that match our target audience and reached out to focus groups where we presented our planned work.
We encouraged them to participate by pointing out the opportunities to learn more about known root causes of test flakiness and to see how other developers and researchers deal with the problem.
Apart from that, no monetary or other incentives were given.

\subsubsection{Global Survey}%
\label{sec:global_survey}

To retrieve a global sample of our target audience, we used Prolific~\cite{Palan2018}, an online service for recruiting subjects for scientific experiments.
We employed a multi-stage filtering process to ensure that the participants match our target audience:
First, we filtered for full-time and part-time employed individuals stating that they have programming experience and are proficient in at least one software development technique from Prolific's 17-element-long list.
Furthermore, we only considered participants with a \SI{100}{\percent} approval rating, meaning that they participated honestly in multiple other studies.
These pre-defined options provided by Prolific resulted in a sample of about \num{7200} out of \num{260000} recently active Prolific members.

Since we are specifically looking for professional software programmers and testers, and Prolific does not provide a suitable filtering option for the participants' occupation, we conducted a prescreening study recruiting \preSurveyNumOfSubmissionsUnique individuals who matched the above demographic options, asking them if they \enquote*{professionally develop or test software}, \enquote*{currently work on a software project}, and \enquote*{write code or tests in their project}.
Like all Prolific studies, the survey was simultaneously advertised to all members matching the demographic filter until the desired number of submissions had been reached, meaning that participants were sampled on a first-come, first-serve principle.
Of the \preSurveyNumOfSubmissionsTrippleYes prescreening study participants who answered all three filter questions in the affirmative, \globalNumNotFullyIncompleteSubmissions participated in our main survey.
Participants received an above-average financial reward (based on Prolific's recommendation).

\subsection{Survey Design}%
\label{sec:overall_survey_design}

\begin{table}[]
\centering
\caption{Structure of the questionnaire.}
\label{tab:questionnaire_structure}
\begin{tabularx}{\columnwidth}{%
    r%
    l%
    X%
    @{}
    p{0cm} %
    r%
    @{\hspace{\tabcolsep}}
    c%
    @{\hspace{\tabcolsep}}
    l%
}
\toprule
    \multicolumn{2}{l}{Question} &
    Type~~~(* allows\newline additional free text) &
    \multicolumn{4}{p{2.2cm}}{Answering Options ($\times$ Statements)}   \\
\midrule
\multicolumn{5}{l}{\textit{Demographic}} \\
\midrule
 1. & Platforms         & multiple-choice*     & & 12 & & \\
 2. & Software Types    & multiple-choice*     & & 20 & & \\
 3. & Code Size         & single-choice scale  & & 6  & & \\
 4. & Team Size         & single-choice scale  & & 6  & & \\
 5. & Test Levels       & multiple-choice*     & & 7  & & \\
 6. & Testing Practices & multiple-choice*     & & 8  & & \\
\midrule
\multicolumn{5}{l}{\textit{Experience with Flakiness}} \\
\midrule
 7.  & Prevalence            & single-choice scale & & 6 & &             \\
 8.  & Severity              & single-choice scale & & 4 & &             \\
 9.  & Root Causes           & Likert scale*      & & 5 & $\times$ & 21 \\
 10. & Consequences          & Likert scale*      & & 5 & $\times$ & \phantom{0}6  \\
 11. & Mitigation Strategies & Likert scale*      & & 5 & $\times$ & 12 \\
 12. & Wishes                & free text           & &   & &             \\
\bottomrule
\end{tabularx}
\end{table}

The survey\footnote{\figshareURL} starts with a short introduction to the issue of test flakiness.
To establish a common understanding of what flaky tests are, it defines them as \enquote*{\textit{Tests that behave non-deterministically, so they sometimes pass, sometimes fail, even though there were no changes in the code they test}}, followed by two examples taken from a previous study~\cite{Gruber2021}.

The questionnaire consists of 12 questions, of which the first six are demographic questions, and the other six are related to the participant's view on flakiness (\cref{tab:questionnaire_structure}).
The latter are directly linked to the research questions.
The survey allows participants to skip questions for which they have no opinion to avoid demotivating them or receiving random answers.
To refine our survey, we piloted it on eight developers from outside and inside BMW, who were not included in the final survey, and revised based on their feedback.

\subsubsection{Demographic Questions}

We avoid asking about specific languages or frameworks as these come in an overwhelming number and follow current trends and are therefore subject to frequent changes.
Instead, our survey starts with two questions asking participants about (1) the platforms they are developing for (such as \textit{Desktop}, \textit{Mobile}, or \textit{IoT}) and (2) the types of software they are developing (such as \textit{Entertainment} or \textit{Security}).
We believe that these statements offer a proper abstraction that is more stable over time while still allowing the inference of possibly used technology stacks.
Both questions are taken from Jetbrains's \textit{State of Developer Ecosystem 2020}~\cite{jetbrains2020} survey.
Our survey provides the same answering options as the Jetbrains study plus one that has been identified while piloting the questionnaire (software type \textit{Education}), as well as automotive-specific options:
The platforms \textit{Classic} and \textit{Adaptive} AUTOSAR\footnote{Automotive Open System Architecture (\url{https://www.autosar.org/})} are the two most common architectures of electronic control units (ECUs) including software running in cars.
The Classic platform is used for computationally less demanding purposes like climate control but also for safety-relevant tasks such as airbag release and stability control systems due to its real-time computing capabilities.
The Adaptive platform has more computing power and is, among others, used to implement automated driving functionalities.
Furthermore, our survey offers three automotive-specific types of software: \textit{Chassis} (e.g., anti-lock breaks), \textit{Drive Train} (e.g., engine control), \textit{Automated Driving and Driver Assistance} (e.g., cruise control and lane-keeping); and adds \textit{Infotainment} (e.g., navigation) to the existing category \textit{Entertainment}.

To estimate the size of our participants' operations, they are asked (3) to specify the number of lines of source code of their main product or project as well as (4) how many developers are working on it.
Both are single-choice questions where the answering options follow a logarithmic scale.
Lastly, our survey asks about (5) the usage of different kinds of tests like \textit{unit} or \textit{integration tests} and (6) testing practices such as \textit{regression testing} or \textit{fuzzing}.
The answering options consist of commonly established testing levels and practices as well as more automotive-specific choices like \textit{hardware-in-the-loop tests}.
For multiple-choice questions, participants can add further options using free text.

\subsubsection{\prevalenceRQ: Prevalence \& Severity}
\label{sec:setup_rq1}

Questions (7) and (8) ask the developers how often they experience flaky failures and how problematic they perceive them.
The prevalence should be specified on a 6-point scale ranging from \textit{never} to \textit{multiple times a day}.
To measure the severity of flakiness, participants should state if it poses to them a \textit{non-existing}, \textit{minor}, \textit{moderate}, or \textit{serious problem}.
These scales match the ones used by Eck et al.~\cite{Eck2019}, allowing us to compare our findings to theirs.

\subsubsection{\rootCauseRQ: Root Causes}
\label{sec:setup_rq3}

To investigate where flakiness comes from, the survey presents the participants (9) with 21 statements describing different root causes of flaky test behavior and asks them to specify how often they experience each on a qualitative 5-point Likert scale (\textit{never} to \textit{always}), following the template by Kasunic~\cite{Kasunic2005}.
They can also add options to the list via a free text field.

We derived the statements from 15 root causes identified by prior studies~\cite{Luo2014, Eck2019, Gruber2021}, splitting \textit{networking} into local bad socket management and remote connection failure~\cite{Luo2014}, and \textit{order dependency} into victims and brittles~\cite{Shi2019}.
These distinctions make our survey both more precise and more easily understandable for the participants.
Furthermore, we added three root causes, which were mentioned to us while piloting the survey, namely reading values from \textit{uninitialized variables}, \textit{differences between compilers} (like undefined behavior or optimizations), and the usage of \textit{random, non-seeded values for testing}.
In addition, participants can also indicate that they do not know what causes their tests to become flaky.
To ensure that the statements are easily understandable, each of them is formulated in a concise sentence, e.g., \enquote*{Prior test did not clean up properly after running} representing the root cause \textit{order dependent victim}.

\subsubsection{\consequencesRQ: Consequences}
\label{sec:setup_rq2}

Using another matrix of Likert scale questions (10), the survey addresses the consequences of flakiness by presenting the participants with six statements describing possible effects of having flaky tests in their test suite, that we derived from literature.
First, developers are asked how often they see 1) failing tests being rerun without being analyzed or 2) passing tests being rerun suspecting hidden bugs.
These two statements investigate the loss of trust in testing.
Second, they are asked how often 3) time or 4) computational resources have to be spent for rerunning, debugging, or fixing flaky tests.
Lastly, the developers should rate how often 5) pull requests cannot be merged and 6) releases are delayed due to test flakiness.
Apart from rating the prepared statements, participants are encouraged to mention further consequences via a free text field.

\subsubsection{\mitigationsRQ: Mitigation Strategies}
\label{sec:setup_rq4}

Since we do not expect the participants to be familiar with specific tools, they have to (11)~rate their usage of the different types of mitigation strategies outlined in \cref{sec:background_mitigations} on a Likert scale and specify any other strategies they apply via free text options.

\subsubsection{\wishesRQ: Wishes}%
\label{sec:setup_rq5}

In the last question, participants are asked (12) what information or tools would be most helpful to them in order to better handle flaky tests.
Unlike the other questions, there are no pre-defined answers, only a free text field.
Through this question, we hope to identify possible directions for future research in order for it to generate practical value.

\subsection{Processing Answers}%
\label{sec:coding_free_text_answers}

\subsubsection{Consolidating Likert Scale Answers}
In order to concisely depict the responses to each Likert scale statement in the limited space of this paper, we transformed the data into integer numbers and calculated their mean using the following mapping: Never~$\rightarrow$~0, Rarely~$\rightarrow$~1, Sometimes~$\rightarrow$~2, Very~Often~$\rightarrow$~3, Always~$\rightarrow$~4.
This allowed us to compress the answers for each statement to a single number.

\subsubsection{Coding Free Text Answers}%
For multiple-choice and Likert scale questions our survey allows additional free text answers.
Another source of free text answers is the last question on the developers' wishes.
Following existing guidelines for conducting surveys in software engineering~\cite{Linaker2015}, we coded these responses using a quantitative content analysis:
As an initial set of codes, we used the pre-defined answering options; for the wishes question we used the categories of the mitigation strategies.
If necessary, we modified this schema by adding new codes, in which case we re-applied the new schema to all data already coded.
The coding was individually executed by both the first author and another software testing researcher.
In case of disagreements, which appeared in 28 out of 265 cases, these were resolved via an in-depth discussion.
Note that one response can have multiple codes assigned to it since some mention multiple different aspects.

\subsection{Relating Answers}%
\label{sec:correlating_answers}

To measure the impact of demographic variables on how developers experience flakiness, we used ordinal logistic regression~\cite{bruin2011}:
We trained a model to predict each response that the participants gave about their experience with flakiness (like prevalence, severity, or any Likert scale statement), based on their answers to the demographic questions.
For each demographic variable, we then looked at its coefficient and confidence interval to see if it constitutes a good predictor.
We considered a connection to be significant if the \SI{95}{\percent} confidence interval does not cross the line of no effect, which is zero for the coefficient.
To measure the strength and direction of a connection, we calculated its odds ratio (OR), which is the exponentiated coefficient.
An odds ratio of \num{1.5} between the variable \textit{platform: Desktop} and \textit{prevalence} would for example mean, that for participants developing desktop applications, the odds of being more likely to experience flaky failures are \num{1.5} times that of participants not developing desktop applications, holding constant all other variables.
An odds ratio above one indicates a positive relation, whereas an odds ratio below one indicates a negative influence on the target variable.

A possible issue with this method is that demographic variables might be correlated, leading to multicollinearity, which reduces the precision of the estimated coefficients and their significance~\cite{Frost}.
To test our data for multicollinearity, we calculated the variance inflation factor (VIF) for each demographic variable and removed variables with elevated values.
This was the case for the size of the participants' code base and the size of their teams.
Excluding these variables also allowed us to use our entire dataset of \bothNumNotFullyIncompleteSubmissions responses since three participants did not answer these two questions.

\subsection{Threats to Validity}%
\label{sec:threats_to_validity}

\subsubsection{External Validity}%
\label{sec:external_validity}

The \globalNumNotFullyIncompleteSubmissions global participants hired via Prolific~\cite{Palan2018} showed typical demographic attributes for the target population (\cref{sec:results_demographics}), however, the mere fact that a developer is active on Prolific might itself carry a bias.
Similarly, for the \automotiveNumNotFullyIncompleteSubmissions automotive participants, their answers might not be representative of the entirety of automotive software engineers.
Furthermore, as we found significant differences between developers working in various domains, their combined results will be influenced by the distribution of the regarded population.
For this reason, we conducted our survey on two different sample populations and report their results separately.

\subsubsection{Construct Validity}%
\label{sec:construct_validity}

To select suitable individuals for our global survey, we employed a multi-stage filtering process.
However, since all this information is self-provided and cannot be verified, our selection process is susceptible to false declarations.
Indeed, we excluded three participants who filled out the prescreening study multiple times, giving different responses.
To recruit automotive participants, we used convenience sampling, which might suffer from self-selection or voluntary response bias:
Developers who experience flakiness more often might be more likely to respond to our survey than those who never heard of it.
Therefore, their numbers might be elevated compared to a random sample.

\subsubsection{Internal Validity}%
\label{sec:internal_validity}

Seven out of our 12 questions use multiple-choice answers or statements that should be ranked.
We derived possible answering options from reviewing both white and grey literature and piloting the questionnaire.
While we might have been influenced by our subjective judgement about what constitutes a worthy answering option, this risk is mitigated by the possibility to provide free text answers, which were coded independently by two researchers.
This process might however be influenced by our interpretation of the responses.%

\section{Results}%

\subsection{Number and Quality of Responses}%
\label{sec:number_and_quality_of_responses}

In total, we collected \bothNumNotFullyIncompleteSubmissions at least partially complete submissions, \automotiveNumNotFullyIncompleteSubmissions from automotive participants and \globalNumNotFullyIncompleteSubmissions from global participants.
\bothNumFullyCompleteSubmissions of these are fully complete, meaning that all non-free-text questions have been answered.
For all further analysis, we consider all  \bothNumNotFullyIncompleteSubmissions.

\subsection{Demographics}%
\label{sec:results_demographics}

\begin{table}[]
    \centering
    \caption{%
        Answers to demographic questions.
        Color maps are proportional to values and question specific.
    }%
    \label{tab:demographic_results}
\resizebox{\columnwidth-0.05cm}{!}{%
        \begin{tabular}{lccc}
\toprule
\textit{Platform (PL)}                           & {Global}                                                          & {Auto}                                                          & {Both} \\
\midrule
Desktop                                          & \cellcolor[HTML]{000000} \color[HTML]{f1f1f1} \SI{55}{\percent} & \cellcolor[HTML]{575757} \color[HTML]{f1f1f1} \SI{40}{\percent} & \cellcolor[HTML]{181818} \color[HTML]{f1f1f1} \SI{50}{\percent} \\
Mobile                                           & \cellcolor[HTML]{676767} \color[HTML]{f1f1f1} \SI{37}{\percent} & \cellcolor[HTML]{d6d6d6} \color[HTML]{000000} \SI{15}{\percent} & \cellcolor[HTML]{898989} \color[HTML]{f1f1f1} \SI{30}{\percent} \\
Web (Back-end)                                   & \cellcolor[HTML]{2e2e2e} \color[HTML]{f1f1f1} \SI{47}{\percent} & \cellcolor[HTML]{b9b9b9} \color[HTML]{000000} \SI{22}{\percent} & \cellcolor[HTML]{5d5d5d} \color[HTML]{f1f1f1} \SI{39}{\percent} \\
Web (Front-end)                                  & \cellcolor[HTML]{0c0c0c} \color[HTML]{f1f1f1} \SI{53}{\percent} & \cellcolor[HTML]{d6d6d6} \color[HTML]{000000} \SI{15}{\percent} & \cellcolor[HTML]{525252} \color[HTML]{f1f1f1} \SI{41}{\percent} \\
WebAssembly                                      & \cellcolor[HTML]{fafafa} \color[HTML]{000000} \SI{03}{\percent} & \cellcolor[HTML]{fcfcfc} \color[HTML]{000000} \SI{02}{\percent} & \cellcolor[HTML]{fbfbfb} \color[HTML]{000000} \SI{02}{\percent} \\
Server / Infrastructure                          & \cellcolor[HTML]{c4c4c4} \color[HTML]{000000} \SI{19}{\percent} & \cellcolor[HTML]{9d9d9d} \color[HTML]{f1f1f1} \SI{26}{\percent} & \cellcolor[HTML]{bababa} \color[HTML]{000000} \SI{21}{\percent} \\
IoT / Embedded                                   & \cellcolor[HTML]{e7e7e7} \color[HTML]{000000} \SI{10}{\percent} & \cellcolor[HTML]{323232} \color[HTML]{f1f1f1} \SI{46}{\percent} & \cellcolor[HTML]{bdbdbd} \color[HTML]{000000} \SI{21}{\percent} \\
Classic AUTOSAR ECU (CAS)                        & \cellcolor[HTML]{fefefe} \color[HTML]{000000} \SI{01}{\percent} & \cellcolor[HTML]{c6c6c6} \color[HTML]{000000} \SI{19}{\percent} & \cellcolor[HTML]{f2f2f2} \color[HTML]{000000} \SI{06}{\percent} \\
Adaptive AUTOSAR ECU (AAS)                       & \cellcolor[HTML]{ffffff} \color[HTML]{000000} \SI{00}{\percent} & \cellcolor[HTML]{838383} \color[HTML]{f1f1f1} \SI{31}{\percent} & \cellcolor[HTML]{e7e7e7} \color[HTML]{000000} \SI{10}{\percent} \\
Consoles (Xbox / PlayStation / Nintendo etc.)    & \cellcolor[HTML]{f8f8f8} \color[HTML]{000000} \SI{03}{\percent} & \cellcolor[HTML]{fefefe} \color[HTML]{000000} \SI{01}{\percent} & \cellcolor[HTML]{fafafa} \color[HTML]{000000} \SI{03}{\percent} \\
Not decided yet (research / proof of concept)    & \cellcolor[HTML]{f9f9f9} \color[HTML]{000000} \SI{03}{\percent} & \cellcolor[HTML]{fafafa} \color[HTML]{000000} \SI{03}{\percent} & \cellcolor[HTML]{f9f9f9} \color[HTML]{000000} \SI{03}{\percent} \\
I don't develop anything                         & \cellcolor[HTML]{fdfdfd} \color[HTML]{000000} \SI{01}{\percent} & \cellcolor[HTML]{fcfcfc} \color[HTML]{000000} \SI{02}{\percent} & \cellcolor[HTML]{fdfdfd} \color[HTML]{000000} \SI{01}{\percent} \\
Other                                            & \cellcolor[HTML]{fcfcfc} \color[HTML]{000000} \SI{02}{\percent} & \cellcolor[HTML]{e7e7e7} \color[HTML]{000000} \SI{10}{\percent} & \cellcolor[HTML]{f7f7f7} \color[HTML]{000000} \SI{04}{\percent} \\
\midrule
\textit{Software Type (ST)}                      & {}                                                              & {}                                                              & {} \\
\midrule
Automotive Chassis                               & \cellcolor[HTML]{fbfbfb} \color[HTML]{000000} \SI{01}{\percent} & \cellcolor[HTML]{e7e7e7} \color[HTML]{000000} \SI{07}{\percent} & \cellcolor[HTML]{f6f6f6} \color[HTML]{000000} \SI{03}{\percent} \\
Automotive Drive Train                           & \cellcolor[HTML]{fefefe} \color[HTML]{000000} \SI{00}{\percent} & \cellcolor[HTML]{f9f9f9} \color[HTML]{000000} \SI{02}{\percent} & \cellcolor[HTML]{fdfdfd} \color[HTML]{000000} \SI{01}{\percent} \\
Automated Driving, Driver Assistance             & \cellcolor[HTML]{fbfbfb} \color[HTML]{000000} \SI{01}{\percent} & \cellcolor[HTML]{101010} \color[HTML]{f1f1f1} \SI{37}{\percent} & \cellcolor[HTML]{cccccc} \color[HTML]{000000} \SI{12}{\percent} \\
Augmented Reality / Virtual Reality              & \cellcolor[HTML]{f6f6f6} \color[HTML]{000000} \SI{03}{\percent} & \cellcolor[HTML]{ececec} \color[HTML]{000000} \SI{06}{\percent} & \cellcolor[HTML]{f3f3f3} \color[HTML]{000000} \SI{04}{\percent} \\
Business Intell. / Data Science / Machine Learn. & \cellcolor[HTML]{aeaeae} \color[HTML]{000000} \SI{17}{\percent} & \cellcolor[HTML]{c9c9c9} \color[HTML]{000000} \SI{13}{\percent} & \cellcolor[HTML]{b8b8b8} \color[HTML]{000000} \SI{16}{\percent} \\
Blockchain                                       & \cellcolor[HTML]{f3f3f3} \color[HTML]{000000} \SI{04}{\percent} & \cellcolor[HTML]{ffffff} \color[HTML]{000000} \SI{00}{\percent} & \cellcolor[HTML]{f7f7f7} \color[HTML]{000000} \SI{03}{\percent} \\
Database / Data Storage                          & \cellcolor[HTML]{272727} \color[HTML]{f1f1f1} \SI{34}{\percent} & \cellcolor[HTML]{d4d4d4} \color[HTML]{000000} \SI{11}{\percent} & \cellcolor[HTML]{626262} \color[HTML]{f1f1f1} \SI{27}{\percent} \\
Education / Training                             & \cellcolor[HTML]{bbbbbb} \color[HTML]{000000} \SI{15}{\percent} & \cellcolor[HTML]{f9f9f9} \color[HTML]{000000} \SI{02}{\percent} & \cellcolor[HTML]{d3d3d3} \color[HTML]{000000} \SI{11}{\percent} \\
Entertainment / Infotainment                     & \cellcolor[HTML]{cbcbcb} \color[HTML]{000000} \SI{12}{\percent} & \cellcolor[HTML]{7b7b7b} \color[HTML]{f1f1f1} \SI{24}{\percent} & \cellcolor[HTML]{b5b5b5} \color[HTML]{000000} \SI{16}{\percent} \\
Fintech (Finance)                                & \cellcolor[HTML]{dbdbdb} \color[HTML]{000000} \SI{09}{\percent} & \cellcolor[HTML]{ffffff} \color[HTML]{000000} \SI{00}{\percent} & \cellcolor[HTML]{e9e9e9} \color[HTML]{000000} \SI{07}{\percent} \\
Games                                            & \cellcolor[HTML]{a7a7a7} \color[HTML]{f1f1f1} \SI{18}{\percent} & \cellcolor[HTML]{f0f0f0} \color[HTML]{000000} \SI{05}{\percent} & \cellcolor[HTML]{c3c3c3} \color[HTML]{000000} \SI{14}{\percent} \\
Hardware                                         & \cellcolor[HTML]{cdcdcd} \color[HTML]{000000} \SI{12}{\percent} & \cellcolor[HTML]{e7e7e7} \color[HTML]{000000} \SI{07}{\percent} & \cellcolor[HTML]{d6d6d6} \color[HTML]{000000} \SI{10}{\percent} \\
Home Automation                                  & \cellcolor[HTML]{ececec} \color[HTML]{000000} \SI{06}{\percent} & \cellcolor[HTML]{f3f3f3} \color[HTML]{000000} \SI{04}{\percent} & \cellcolor[HTML]{eeeeee} \color[HTML]{000000} \SI{05}{\percent} \\
IT Infrastructure                                & \cellcolor[HTML]{aaaaaa} \color[HTML]{f1f1f1} \SI{17}{\percent} & \cellcolor[HTML]{7b7b7b} \color[HTML]{f1f1f1} \SI{24}{\percent} & \cellcolor[HTML]{9c9c9c} \color[HTML]{f1f1f1} \SI{19}{\percent} \\
Libraries / Frameworks                           & \cellcolor[HTML]{e1e1e1} \color[HTML]{000000} \SI{08}{\percent} & \cellcolor[HTML]{676767} \color[HTML]{f1f1f1} \SI{26}{\percent} & \cellcolor[HTML]{c3c3c3} \color[HTML]{000000} \SI{14}{\percent} \\
Programming Tools                                & \cellcolor[HTML]{cbcbcb} \color[HTML]{000000} \SI{12}{\percent} & \cellcolor[HTML]{979797} \color[HTML]{f1f1f1} \SI{20}{\percent} & \cellcolor[HTML]{bebebe} \color[HTML]{000000} \SI{15}{\percent} \\
Security                                         & \cellcolor[HTML]{dddddd} \color[HTML]{000000} \SI{09}{\percent} & \cellcolor[HTML]{ececec} \color[HTML]{000000} \SI{06}{\percent} & \cellcolor[HTML]{e1e1e1} \color[HTML]{000000} \SI{08}{\percent} \\
System Software (e.g. OS driver)                 & \cellcolor[HTML]{ececec} \color[HTML]{000000} \SI{06}{\percent} & \cellcolor[HTML]{cecece} \color[HTML]{000000} \SI{12}{\percent} & \cellcolor[HTML]{e3e3e3} \color[HTML]{000000} \SI{08}{\percent} \\
Utilities (small apps for small tasks)           & \cellcolor[HTML]{545454} \color[HTML]{f1f1f1} \SI{29}{\percent} & \cellcolor[HTML]{676767} \color[HTML]{f1f1f1} \SI{26}{\percent} & \cellcolor[HTML]{5b5b5b} \color[HTML]{f1f1f1} \SI{28}{\percent} \\
Websites                                         & \cellcolor[HTML]{000000} \color[HTML]{f1f1f1} \SI{39}{\percent} & \cellcolor[HTML]{d4d4d4} \color[HTML]{000000} \SI{11}{\percent} & \cellcolor[HTML]{474747} \color[HTML]{f1f1f1} \SI{31}{\percent} \\
Other                                            & \cellcolor[HTML]{ededed} \color[HTML]{000000} \SI{06}{\percent} & \cellcolor[HTML]{d4d4d4} \color[HTML]{000000} \SI{11}{\percent} & \cellcolor[HTML]{e6e6e6} \color[HTML]{000000} \SI{07}{\percent} \\
\midrule
\textit{Code Size}                               & {}                                                              & {}                                                              & {} \\
\midrule
$<$ 1k                                           & \cellcolor[HTML]{bebebe} \color[HTML]{000000} \SI{12}{\percent} & \cellcolor[HTML]{fcfcfc} \color[HTML]{000000} \SI{01}{\percent} & \cellcolor[HTML]{d6d6d6} \color[HTML]{000000} \SI{09}{\percent} \\
1k - 10k                                         & \cellcolor[HTML]{030303} \color[HTML]{f1f1f1} \SI{32}{\percent} & \cellcolor[HTML]{dbdbdb} \color[HTML]{000000} \SI{08}{\percent} & \cellcolor[HTML]{4e4e4e} \color[HTML]{f1f1f1} \SI{25}{\percent} \\
10k - 100k                                       & \cellcolor[HTML]{000000} \color[HTML]{f1f1f1} \SI{33}{\percent} & \cellcolor[HTML]{464646} \color[HTML]{f1f1f1} \SI{25}{\percent} & \cellcolor[HTML]{141414} \color[HTML]{f1f1f1} \SI{30}{\percent} \\
100k - 1M                                        & \cellcolor[HTML]{9e9e9e} \color[HTML]{f1f1f1} \SI{15}{\percent} & \cellcolor[HTML]{020202} \color[HTML]{f1f1f1} \SI{32}{\percent} & \cellcolor[HTML]{717171} \color[HTML]{f1f1f1} \SI{21}{\percent} \\
1M - 10M                                         & \cellcolor[HTML]{ececec} \color[HTML]{000000} \SI{05}{\percent} & \cellcolor[HTML]{939393} \color[HTML]{f1f1f1} \SI{17}{\percent} & \cellcolor[HTML]{d8d8d8} \color[HTML]{000000} \SI{08}{\percent} \\
$>$ 10M                                          & \cellcolor[HTML]{f4f4f4} \color[HTML]{000000} \SI{03}{\percent} & \cellcolor[HTML]{a5a5a5} \color[HTML]{f1f1f1} \SI{15}{\percent} & \cellcolor[HTML]{e2e2e2} \color[HTML]{000000} \SI{07}{\percent} \\
no answer                                        & \cellcolor[HTML]{ffffff} \color[HTML]{000000} \SI{00}{\percent} & \cellcolor[HTML]{f8f8f8} \color[HTML]{000000} \SI{02}{\percent} & \cellcolor[HTML]{fdfdfd} \color[HTML]{000000} \SI{01}{\percent} \\
\midrule
\textit{Team Size}                               & {}                                                              & {}                                                              & {} \\
\midrule
Just me                                          & \cellcolor[HTML]{919191} \color[HTML]{f1f1f1} \SI{23}{\percent} & \cellcolor[HTML]{f2f2f2} \color[HTML]{000000} \SI{05}{\percent} & \cellcolor[HTML]{b6b6b6} \color[HTML]{000000} \SI{18}{\percent} \\
2 - 5                                            & \cellcolor[HTML]{000000} \color[HTML]{f1f1f1} \SI{45}{\percent} & \cellcolor[HTML]{949494} \color[HTML]{f1f1f1} \SI{23}{\percent} & \cellcolor[HTML]{2e2e2e} \color[HTML]{f1f1f1} \SI{38}{\percent} \\
6 - 10                                           & \cellcolor[HTML]{c5c5c5} \color[HTML]{000000} \SI{15}{\percent} & \cellcolor[HTML]{ebebeb} \color[HTML]{000000} \SI{07}{\percent} & \cellcolor[HTML]{d3d3d3} \color[HTML]{000000} \SI{13}{\percent} \\
11 - 20                                          & \cellcolor[HTML]{e6e6e6} \color[HTML]{000000} \SI{08}{\percent} & \cellcolor[HTML]{c3c3c3} \color[HTML]{000000} \SI{16}{\percent} & \cellcolor[HTML]{dcdcdc} \color[HTML]{000000} \SI{10}{\percent} \\
21 - 100                                         & \cellcolor[HTML]{ebebeb} \color[HTML]{000000} \SI{07}{\percent} & \cellcolor[HTML]{d1d1d1} \color[HTML]{000000} \SI{13}{\percent} & \cellcolor[HTML]{e4e4e4} \color[HTML]{000000} \SI{09}{\percent} \\
$>$ 100                                          & \cellcolor[HTML]{f9f9f9} \color[HTML]{000000} \SI{02}{\percent} & \cellcolor[HTML]{3a3a3a} \color[HTML]{f1f1f1} \SI{36}{\percent} & \cellcolor[HTML]{d3d3d3} \color[HTML]{000000} \SI{13}{\percent} \\
no answer                                        & \cellcolor[HTML]{ffffff} \color[HTML]{000000} \SI{00}{\percent} & \cellcolor[HTML]{fdfdfd} \color[HTML]{000000} \SI{01}{\percent} & \cellcolor[HTML]{ffffff} \color[HTML]{000000} \SI{00}{\percent} \\
\midrule
\textit{Test Level (TL)}                         & {}                                                              & {}                                                              & {} \\
\midrule
Unit Tests                                       & \cellcolor[HTML]{717171} \color[HTML]{f1f1f1} \SI{61}{\percent} & \cellcolor[HTML]{000000} \color[HTML]{f1f1f1} \SI{94}{\percent} & \cellcolor[HTML]{535353} \color[HTML]{f1f1f1} \SI{71}{\percent} \\
Component Tests                                  & \cellcolor[HTML]{d5d5d5} \color[HTML]{000000} \SI{27}{\percent} & \cellcolor[HTML]{797979} \color[HTML]{f1f1f1} \SI{58}{\percent} & \cellcolor[HTML]{bebebe} \color[HTML]{000000} \SI{37}{\percent} \\
Integration Tests                                & \cellcolor[HTML]{8c8c8c} \color[HTML]{f1f1f1} \SI{52}{\percent} & \cellcolor[HTML]{232323} \color[HTML]{f1f1f1} \SI{83}{\percent} & \cellcolor[HTML]{6e6e6e} \color[HTML]{f1f1f1} \SI{61}{\percent} \\
System Tests                                     & \cellcolor[HTML]{b5b5b5} \color[HTML]{000000} \SI{39}{\percent} & \cellcolor[HTML]{8f8f8f} \color[HTML]{f1f1f1} \SI{51}{\percent} & \cellcolor[HTML]{a9a9a9} \color[HTML]{f1f1f1} \SI{43}{\percent} \\
Software in the loop Tests (SiL)                 & \cellcolor[HTML]{f0f0f0} \color[HTML]{000000} \SI{15}{\percent} & \cellcolor[HTML]{d7d7d7} \color[HTML]{000000} \SI{26}{\percent} & \cellcolor[HTML]{e9e9e9} \color[HTML]{000000} \SI{18}{\percent} \\
Platform in the loop Tests (PiL)                 & \cellcolor[HTML]{f9f9f9} \color[HTML]{000000} \SI{08}{\percent} & \cellcolor[HTML]{f5f5f5} \color[HTML]{000000} \SI{11}{\percent} & \cellcolor[HTML]{f7f7f7} \color[HTML]{000000} \SI{09}{\percent} \\
Hardware in the loop Tests (HiL)                 & \cellcolor[HTML]{fcfcfc} \color[HTML]{000000} \SI{05}{\percent} & \cellcolor[HTML]{e7e7e7} \color[HTML]{000000} \SI{19}{\percent} & \cellcolor[HTML]{f7f7f7} \color[HTML]{000000} \SI{09}{\percent} \\
Other                                            & \cellcolor[HTML]{fcfcfc} \color[HTML]{000000} \SI{05}{\percent} & \cellcolor[HTML]{ffffff} \color[HTML]{000000} \SI{03}{\percent} & \cellcolor[HTML]{fdfdfd} \color[HTML]{000000} \SI{04}{\percent} \\
\midrule
\textit{Testing Practices (TP)}                  & {}                                                              & {}                                                              & {} \\
\midrule
Manual Testing                                   & \cellcolor[HTML]{212121} \color[HTML]{f1f1f1} \SI{85}{\percent} & \cellcolor[HTML]{696969} \color[HTML]{f1f1f1} \SI{64}{\percent} & \cellcolor[HTML]{393939} \color[HTML]{f1f1f1} \SI{79}{\percent} \\
Automated Testing                                & \cellcolor[HTML]{6f6f6f} \color[HTML]{f1f1f1} \SI{61}{\percent} & \cellcolor[HTML]{171717} \color[HTML]{f1f1f1} \SI{88}{\percent} & \cellcolor[HTML]{585858} \color[HTML]{f1f1f1} \SI{70}{\percent} \\
Continuous Integration                           & \cellcolor[HTML]{c3c3c3} \color[HTML]{000000} \SI{33}{\percent} & \cellcolor[HTML]{000000} \color[HTML]{f1f1f1} \SI{96}{\percent} & \cellcolor[HTML]{898989} \color[HTML]{f1f1f1} \SI{53}{\percent} \\
Regression Testing                               & \cellcolor[HTML]{d6d6d6} \color[HTML]{000000} \SI{25}{\percent} & \cellcolor[HTML]{737373} \color[HTML]{f1f1f1} \SI{60}{\percent} & \cellcolor[HTML]{bebebe} \color[HTML]{000000} \SI{36}{\percent} \\
Fuzzing                                          & \cellcolor[HTML]{fafafa} \color[HTML]{000000} \SI{04}{\percent} & \cellcolor[HTML]{f0f0f0} \color[HTML]{000000} \SI{12}{\percent} & \cellcolor[HTML]{f7f7f7} \color[HTML]{000000} \SI{07}{\percent} \\
Property-based Testing                           & \cellcolor[HTML]{f8f8f8} \color[HTML]{000000} \SI{06}{\percent} & \cellcolor[HTML]{fdfdfd} \color[HTML]{000000} \SI{02}{\percent} & \cellcolor[HTML]{fafafa} \color[HTML]{000000} \SI{04}{\percent} \\
Parameterized Tests                              & \cellcolor[HTML]{eeeeee} \color[HTML]{000000} \SI{13}{\percent} & \cellcolor[HTML]{8e8e8e} \color[HTML]{f1f1f1} \SI{51}{\percent} & \cellcolor[HTML]{d8d8d8} \color[HTML]{000000} \SI{24}{\percent} \\
Test Generation                                  & \cellcolor[HTML]{f6f6f6} \color[HTML]{000000} \SI{07}{\percent} & \cellcolor[HTML]{efefef} \color[HTML]{000000} \SI{13}{\percent} & \cellcolor[HTML]{f4f4f4} \color[HTML]{000000} \SI{09}{\percent} \\
Other                                            & \cellcolor[HTML]{ffffff} \color[HTML]{000000} \SI{00}{\percent} & \cellcolor[HTML]{fdfdfd} \color[HTML]{000000} \SI{02}{\percent} & \cellcolor[HTML]{ffffff} \color[HTML]{000000} \SI{01}{\percent} \\
\bottomrule
\end{tabular}

}
\end{table}

Inspecting the responses to the demographic questions of our survey, we see a clear difference between automotive and global participants (\cref{tab:demographic_results}).
Global participants mainly develop websites, databases, and utilities for desktop-, web-, and mobile systems, strongly matching the Jetbrains survey~\cite{jetbrains2020} from where these questions were taken.
Their projects typically encapsulate not more than 100k lines of code and 5 people, which is again similar to the Jetbrains study (\SI{52}{\percent} work in project teams of 2-7 people) and Eck's survey~\cite{Eck2019} (\SI{76}{\percent} work in teams of up to 10 people).
On the other hand, while also working on utilities and libraries, automotive participants develop software for automated driving, IT infrastructure, and infotainment for IoT-, desktop-, and AUTOSAR platforms.
Their operations are larger, comprising more than 100k lines of code and over 20 people on average.
Furthermore, automotive participants make far greater use of automated testing, continuous integration, regression testing, and parametrized testing than global participants.

Through Prolific, we also have access to other demographic data about the global participants, including age, sex, student status, and various nationality and language variables.
We did not record any such data for the automotive survey because we do not use it to answer any of the RQs. However, for the global survey, this data can help us to verify if we achieved our goal of retrieving a global sample of software developers and testers.
In general, participants are rather young, with a majority being under 30 (median \globalAgeMedian, youngest \globalAgeMin, oldest \globalAgeMax), which is also reflected in the high proportion of students (\globalStudentStatusYes).
This is comparable to Jetbrains's \textit{The State of Developer Ecosystem 2020}~\cite{jetbrains2020}, where \SI{24}{\percent} of all participants were (working) students and \SI{63}{\percent} were younger than 30.
In terms of nationality, we see a clear tendency towards the western hemisphere, with \globalNationalityPercentEurope coming from Europe and \globalNationalityPercentNorthAmerica from North America, which is similar to another Prolific-based software engineering survey~\cite{Russo2020} but slightly different from the Jetbrains study, which also features many participants from China and India.
Of our global participants, \globalSexPercentageMale are male, which almost exactly matches the distribution from Russo et al.'s survey~\cite{Russo2020} (\SI{81.4}{\percent}), who argue that this is a comparably high female ratio for the field (\SI{93}{\percent} of participants in Jetbrains's 2021 study were male\footnote{this question was not asked in the 2020 survey.}).
Overall, we see no strong bias towards any specific groups but rather a strong overlap with other representative studies, indicating that we retrieved a diverse, global sample of software developers and testers.

\subsection{\prevalenceRQ: Prevalence \& Severity}
\label{sec:results_rq1}

\begin{figure}[]
    \centering
    \includegraphics[width=\linewidth]{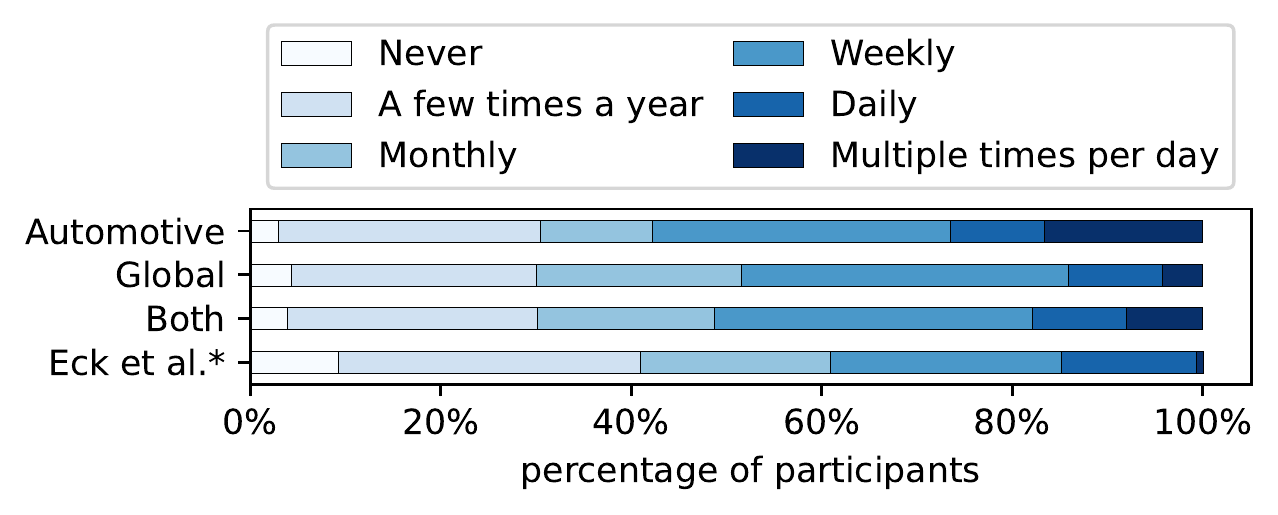}
     \caption{
        How often do you experience flaky failures?\newline
        * Eck et al.\ asked \enquote*{How often do you deal with flaky tests?}
    }
     \label{fig:fl1_frequency}
\end{figure}

\begin{figure}[]
    \centering
    \includegraphics[width=\linewidth]{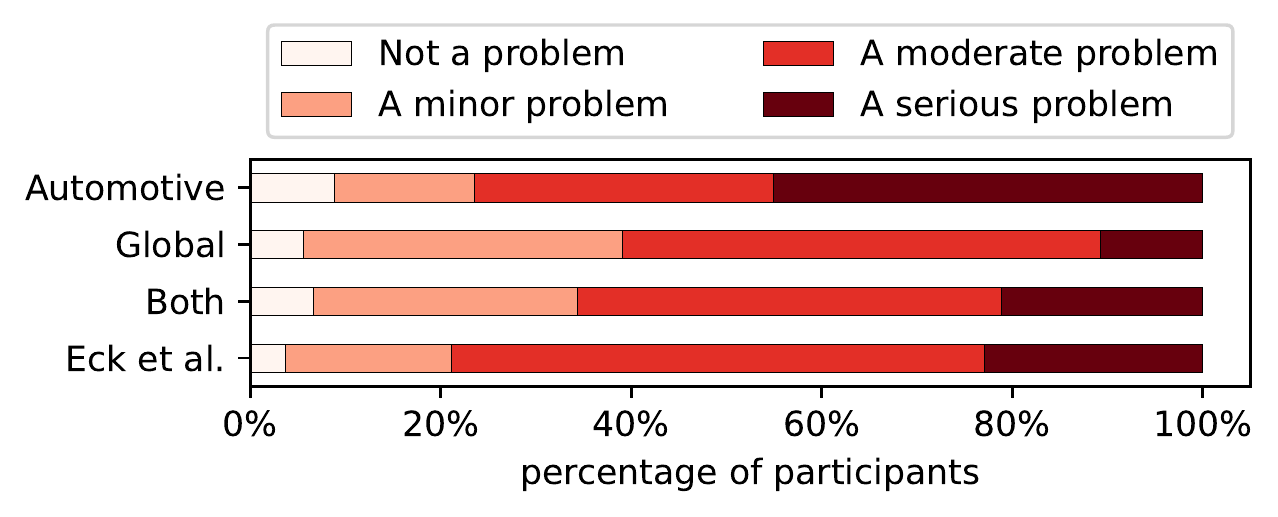}
    \caption{How problematic is flakiness for you?}%
    \label{fig:fl2_severity}
\end{figure}

\Cref{fig:fl1_frequency,fig:fl2_severity} show how frequently participants experience flaky failures and how problematic they perceive flakiness.
Almost every participant has to deal with flaky failures at least a few times a year (only \numBothFreqNever never experience the problem), a majority experiences the issue at least weekly, and \SI{66}{\percent} rate it a moderate or serious issue!

\begin{figure}[]
    \centering
    \includegraphics[width=0.9\linewidth]{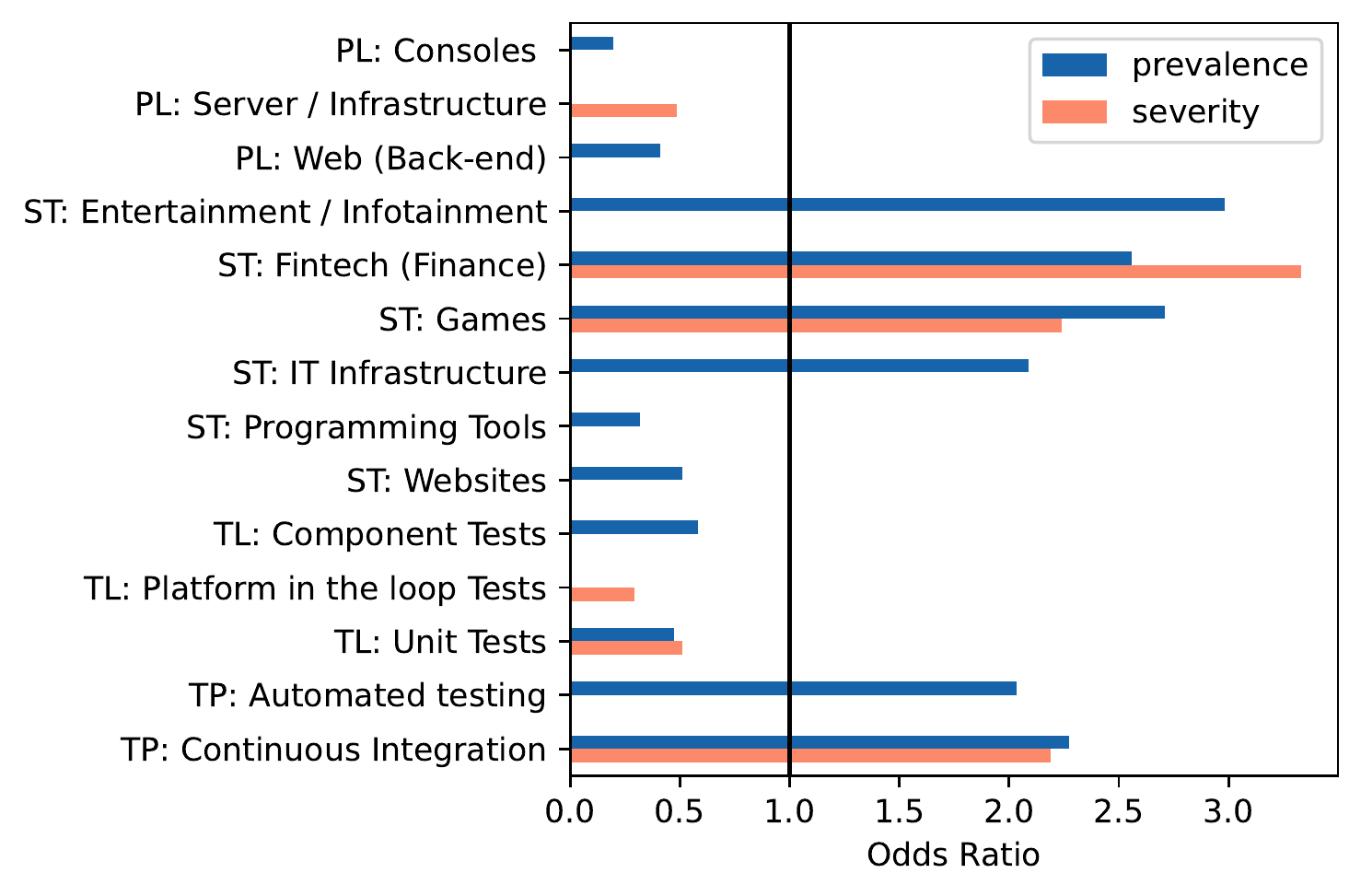}
    \caption{
        Demographic variables with significant impact on prevalence and/or severity of flakiness.
        The black bar separates negative influence ($<1$) from positive influence ($>1$).
    }%
    \label{fig:prevalence_logit}
\end{figure}

We also see a considerable difference between our two participant groups as automotive participants seem to be more affected by flakiness.
Consulting the odds ratios of the demographic variables (\cref{fig:prevalence_logit}) offers an explanation for this observation:
The usage of automated testing and continuous integration (bottom two bars) are both strong positive predictors for the prevalence (and partially severity) of flaky failures, and both testing practices are used by automotive participants at a far higher rate (\cref{tab:demographic_results}).
This confirms other studies reporting flaky failures to be particularly common in continuous integration environments~\cite{Labuschagne2017, Lam2019a} and explains why test flakiness has been an uprising issue for the last decade, where continuous integration has received wide adoption\footnote{both Jenkins and Travis CI celebrate their 10th birthday this year}.
\Cref{fig:prevalence_logit} also shows several unexpected relations: Participants developing websites, particularly back-end applications, seem to experience flaky failures less often despite using continuous integration and automated testing to the same or higher degrees as other participants.
While this offers an additional explanation for why automotive participants are more strongly affected, we would have expected quite the opposite, since one of the most common causes of test flakiness (asynchronous waiting) largely refers to front-end--back-end communication.
Other surprising observations include the strong negative effect of platform \textit{Consoles} and the strong positive effect of software types \textit{Entertainment / Infotainment}, \textit{Fintech}, and \textit{Games}.
However, it has to be noted that these are the least distinctive variables, as only a few participants fall into these groups (min \SI{3}{\percent}, max. \SI{16}{\percent}), particularly compared to continuous integration, which \SI{53}{\percent} of all participants specified, making them more susceptible to the answers of individual participants.

\begin{table}[]
    \centering
    \caption{%
        Means per platform, sorted by prevalence.
        Color-scales are column-specific.
        Only showing platforms with at least 10 participants.
        Rows overlap (share participants) since the platform question allows multiple choice.
    }%
    \label{tab:means_per_platform}
\resizebox{\columnwidth}{!}{%
    \begin{tabular}{lrrrr}
\toprule
{Platform}       & {Prevalence} & {Severity}   & {Code Size}  & {Team Size}  \\
\midrule
{min, max, mean} & {0, 5, 2.43} & {0, 3, 1.80} & {0, 5, 2.15} & {0, 5, 1.92} \\
\midrule
Web (Back-end)          & \cellcolor[HTML]{ffffff} \color[HTML]{000000} 2.13 & \cellcolor[HTML]{ffffff} \color[HTML]{000000} 1.66 & \cellcolor[HTML]{f0f0f0} \color[HTML]{000000} 2.07 & \cellcolor[HTML]{ffffff} \color[HTML]{000000} 1.55  \\
Web (Front-end)         & \cellcolor[HTML]{f6f6f6} \color[HTML]{000000} 2.19 & \cellcolor[HTML]{fefefe} \color[HTML]{000000} 1.67 & \cellcolor[HTML]{f2f2f2} \color[HTML]{000000} 2.05 & \cellcolor[HTML]{ffffff} \color[HTML]{000000} 1.53  \\
Classic AUTOSAR ECU     & \cellcolor[HTML]{d4d4d4} \color[HTML]{000000} 2.33 & \cellcolor[HTML]{6e6e6e} \color[HTML]{f1f1f1} 2.19 & \cellcolor[HTML]{3c3c3c} \color[HTML]{f1f1f1} 3.29 & \cellcolor[HTML]{2f2f2f} \color[HTML]{f1f1f1} 3.95  \\
Server / Infrastructure & \cellcolor[HTML]{a3a3a3} \color[HTML]{f1f1f1} 2.47 & \cellcolor[HTML]{f7f7f7} \color[HTML]{000000} 1.72 & \cellcolor[HTML]{e7e7e7} \color[HTML]{000000} 2.17 & \cellcolor[HTML]{e3e3e3} \color[HTML]{000000} 2.10 \\
Desktop                 & \cellcolor[HTML]{a0a0a0} \color[HTML]{f1f1f1} 2.48 & \cellcolor[HTML]{f4f4f4} \color[HTML]{000000} 1.74 & \cellcolor[HTML]{f3f3f3} \color[HTML]{000000} 2.03 & \cellcolor[HTML]{f7f7f7} \color[HTML]{000000} 1.72  \\
Mobile                  & \cellcolor[HTML]{959595} \color[HTML]{f1f1f1} 2.50 & \cellcolor[HTML]{f9f9f9} \color[HTML]{000000} 1.70 & \cellcolor[HTML]{ffffff} \color[HTML]{000000} 1.85 & \cellcolor[HTML]{fcfcfc} \color[HTML]{000000} 1.61  \\
IoT / Embedded          & \cellcolor[HTML]{585858} \color[HTML]{f1f1f1} 2.67 & \cellcolor[HTML]{8e8e8e} \color[HTML]{f1f1f1} 2.10 & \cellcolor[HTML]{959595} \color[HTML]{f1f1f1} 2.74 & \cellcolor[HTML]{a7a7a7} \color[HTML]{f1f1f1} 2.81  \\
Adaptive AUTOSAR ECU    & \cellcolor[HTML]{000000} \color[HTML]{f1f1f1} 2.88 & \cellcolor[HTML]{000000} \color[HTML]{f1f1f1} 2.48 & \cellcolor[HTML]{000000} \color[HTML]{f1f1f1} 3.62 & \cellcolor[HTML]{000000} \color[HTML]{f1f1f1} 4.39  \\
\bottomrule
\end{tabular}

}
\end{table}

Lastly, we look at the correlation between the two variables prevalence and severity, which is both strong, positive, and significant ($\tau = 0.39$, $p < 0.001$).
While we cover the implications of this observation in the context of RQ3 (\cref{sec:results_rq2_consequences}), there are two noteworthy exceptions (\cref{tab:means_per_platform}):
Developers working on Classic AUTOSAR projects experience flaky failures less often than average while rating flakiness a more serious problem than most other projects.
This observation can be explained by the type of applications running on this platform:
The Classic platform is used for safety-relevant real-time software, such as airbag release or stability control systems.
As a result, these programs have strict requirements towards deterministic behavior, explaining both the rarity and the criticality of flaky failures.
Furthermore, correcting errors in the field can be very costly, especially since software failures might also damage hardware parts.
Participants developing mobile applications show rather the opposite picture:
They experience flaky failures frequently while rating it a less serious issue.
Talking to multiple representatives of this group, we found that this is caused by Android emulators and testing tools, which are plagued by non-determinism.
One participant specifically mentions that they do not wish for any additional tools to tackle flakiness but just for more stable Android testing tools since about \SI{80}{\percent} of their flakiness originates from the test environment, not the tests themselves.
This observation is confirmed by Listfield~\cite{Listfield2017}, who reports that \SI{25}{\percent} of their tests using the Android emulator are flaky (compared to \SI{1.7}{\percent} of all tests) and that the Android Emulator is the only tool for which the increased amount of flakiness cannot be explained by tests simply being larger.

\summary{\prevalenceRQ: Prevalence and Severity of Flakiness}{
    Test flakiness is a common and severe issue, especially among those using continuous integration.
    Domain-specific factors like flaky emulators or safety relevance influence the prevalence and severity of flaky failures.
}

\subsection{\rootCauseRQ: Causes}
\label{sec:results_rq3_causes}

\begin{table}[]
    \centering
    \caption{
        Mean rating per root cause from all participants, ranging from never (0) to always (4).
        For the demographic influences, we excluded groups with less than 10 participants.
    }
    \label{tab:rootCauseTable}
    \resizebox{\columnwidth}{!}{
        \begin{tabular}{lrl}
\toprule
{Root Cause}                & {Both} & {Demographic influences (strongest 3)} \\
\midrule
Concurrency                 & \cellcolor[HTML]{000000} \color[HTML]{f1f1f1} 1.80 & $\myminus$ Research/PoC, $\myminus$ CAS, $\myplus$ Cont. Integr.\\
Order-dependency (victim)   & \cellcolor[HTML]{111111} \color[HTML]{f1f1f1} 1.74 & $\myplus$ Home autom., $\myplus$ Database\\
Async Wait                  & \cellcolor[HTML]{1f1f1f} \color[HTML]{f1f1f1} 1.70 & $\myplus$ Test generation, $\myminus$ System Test\\
Unknown Cause               & \cellcolor[HTML]{393939} \color[HTML]{f1f1f1} 1.63 & $\myminus$ Test gen., $\myplus$ Fintech, $\myplus$ Entertainm.\\
Network (remote)            & \cellcolor[HTML]{464646} \color[HTML]{f1f1f1} 1.60 & $\myminus$ Chassis, $\myplus$ HiL Test, $\myplus$ Games\\
Test case timeout           & \cellcolor[HTML]{545454} \color[HTML]{f1f1f1} 1.56 & $\myplus$ Games, $\myplus$ System SW, $\myplus$ Fintech\\
Test suite timeout          & \cellcolor[HTML]{616161} \color[HTML]{f1f1f1} 1.51 & $\myminus$ Prop. based Test, $\myplus$ System SW, $\myplus$ AR\\
Order-dependency (brittle)  & \cellcolor[HTML]{696969} \color[HTML]{f1f1f1} 1.48 & $\myplus$ Database, $\myminus$ System Test\\
Infrastructure              & \cellcolor[HTML]{828282} \color[HTML]{f1f1f1} 1.40 & $\myplus$ Research/PoC, $\myplus$ Games\\
Uninitialized variables     & \cellcolor[HTML]{aeaeae} \color[HTML]{000000} 1.26 & $\myminus$ Research/PoC, $\myplus$ Hardware\\
Time                        & \cellcolor[HTML]{b0b0b0} \color[HTML]{000000} 1.25 & $\myminus$ CAS, $\myplus$ HiL Test, $\myplus$ Fintech\\
Randomness                  & \cellcolor[HTML]{c4c4c4} \color[HTML]{000000} 1.19 & $\myminus$ Automat. Driving, $\myplus$ Games, $\myplus$ Edu.\\
Floating point              & \cellcolor[HTML]{c9c9c9} \color[HTML]{000000} 1.16 & $\myplus$ Games, $\myplus$ Fintech, $\myplus$ PiL Test\\
Too restrictive range       & \cellcolor[HTML]{cfcfcf} \color[HTML]{000000} 1.13 & $\myminus$ CAS, $\myplus$ Fintech, $\myplus$ Games\\
Resource Leaks              & \cellcolor[HTML]{dadada} \color[HTML]{000000} 1.09 & $\myplus$ Games, $\myminus$ Prog. Tools, $\myplus$ Sever/Infr.\\
Compiler differences        & \cellcolor[HTML]{dbdbdb} \color[HTML]{000000} 1.09 & $\myplus$ Fintech, $\myplus$ Fuzzing, $\myplus$ Hardware \\
Unordered collection        & \cellcolor[HTML]{e0e0e0} \color[HTML]{000000} 1.06 & $\myplus$ Hardware, $\myplus$ Education\\
Platform dependency         & \cellcolor[HTML]{e0e0e0} \color[HTML]{000000} 1.06 & $\myplus$ Fuzzing, $\myplus$ PiL Test, $\myplus$ Fintech \\
Non-Seeded Testing          & \cellcolor[HTML]{f3f3f3} \color[HTML]{000000} 0.96 & $\myplus$ Fuzzing, $\myplus$ Hardware, $\myplus$ Comp. Test\\
Network (local)             & \cellcolor[HTML]{f6f6f6} \color[HTML]{000000} 0.93 & $\myplus$ Games, $\myplus$ Hardware, $\myminus$ Manual Test\\
IO (garbage collector)      & \cellcolor[HTML]{ffffff} \color[HTML]{000000} 0.86 & $\myplus$ Fintech, $\myplus$ Fuzzing, $\myplus$ PiL Test\\
\bottomrule
\end{tabular}

    }
\end{table}

\cref{tab:rootCauseTable} depicts our results regarding the causes of flakiness.
In contrast to the previous question, there is strong agreement between global and automotive participants:
Both rank concurrency as the most prevalent cause of flakiness and order-dependent victims, asynchronous waiting, and unknown causes among the top five.
For each type of root cause, \cref{tab:rootCauseTable} lists the three most important demographic factors determined using logistic regression, denoting whether the factor leads to higher or lower prevalence. We observe multiple phenomena of which we highlight three trends we found most distinctive:
\begin{itemize}
\item First, concurrency-related issues are less common in projects that have not decided on a target platform, such as research projects or proofs of concept.
This might be the case because such software has little use for concurrency, as it is neither relevant for performance optimization through parallelization nor for multi-user interaction.
Developers targeting Classic AUTOSAR ECUs---being quite the opposite of a proof of concept---also experience fewer issues with concurrency, however, this is the case because concurrent processes are meticulously controlled using time slices, avoiding unforeseen behavior (see also root cause \textit{time}).
\item Second, we see that developers working on databases have issues with order-dependencies, which is intuitive since these are main carriers of global states that have to be properly reset after every test execution.
\item Lastly, the two new categories \textit{Uninitialized variables} and \textit{Compiler differences}, which were pointed out to us by automotive developers during the survey pilot, are experienced by hardware developers and are also common with global participants.
This is worth mentioning, especially since these causes only exist in certain languages.
\end{itemize}
We also received 27 free text answers, largely from automotive participants (20x), who mention infrastructure issues (9x), mostly concerning CI (6x), for example, failing to pull docker images, as well as flaky behavior of the device under test (4x).
Both are indicative of the type of software they develop, which is largely embedded programs requiring specialized hardware setups. These may result in difficult to control dependencies between hardware and environment~\cite{Strandberg2020}.
Global participants frequently mention old code and undocumented changes (7x), for example, to remote APIs, causing tests to be flaky.

\summary{\rootCauseRQ: Root Causes of Flakiness}{
    While causes vary for different kinds of software and tests, concurrency and order-dependencies are the most common causes of test flakiness.
    Uninitialized variables, differences between compilers, and unexpected API behavior are further previously unmentioned but not uncommon reasons for flaky tests.
    However, in many cases, the root causes remain unknown.
}

\subsection{\consequencesRQ: Consequences}
\label{sec:results_rq2_consequences}

\begin{table}[]
    \centering
    \caption{
        Mean rating per consequence; never (0) - always (4).
    }
    \label{tab:consequencesTable}
\begin{tabular}{lccc}
\toprule
{Consequences}                      & {Global}                                           & {Auto}                                             & {Both} \\
\midrule
Wasting developer time              & \cellcolor[HTML]{818181} \color[HTML]{f1f1f1} 2.32 & \cellcolor[HTML]{232323} \color[HTML]{f1f1f1} 2.73 & \cellcolor[HTML]{676767} \color[HTML]{f1f1f1} 2.44 \\
Rerun failures without analyzing    & \cellcolor[HTML]{d7d7d7} \color[HTML]{000000} 1.89 & \cellcolor[HTML]{555555} \color[HTML]{f1f1f1} 2.54 & \cellcolor[HTML]{b4b4b4} \color[HTML]{000000} 2.08 \\
Merging pull requests is harder               & \cellcolor[HTML]{ededed} \color[HTML]{000000} 1.73 & \cellcolor[HTML]{000000} \color[HTML]{f1f1f1} 2.89 & \cellcolor[HTML]{b5b5b5} \color[HTML]{000000} 2.08 \\
Wasting computational resources     & \cellcolor[HTML]{f6f6f6} \color[HTML]{000000} 1.65 & \cellcolor[HTML]{545454} \color[HTML]{f1f1f1} 2.54 & \cellcolor[HTML]{d4d4d4} \color[HTML]{000000} 1.91 \\
Rerun passes suspecting hidden bugs & \cellcolor[HTML]{c5c5c5} \color[HTML]{000000} 2.00 & \cellcolor[HTML]{ffffff} \color[HTML]{000000} 1.54 & \cellcolor[HTML]{dcdcdc} \color[HTML]{000000} 1.86 \\
Delayed releases                    & \cellcolor[HTML]{ededed} \color[HTML]{000000} 1.73 & \cellcolor[HTML]{dcdcdc} \color[HTML]{000000} 1.86 & \cellcolor[HTML]{e8e8e8} \color[HTML]{000000} 1.77 \\
\bottomrule
\end{tabular}

\end{table}

\cref{tab:consequencesTable} shows the average ratings to the six statements from question nine, \enquote*{Which negative effects of flaky tests have you experienced?}
Wasting developer time is perceived as the most severe consequence of test flakiness.
Nevertheless, other impediments are also present in everyday business:
For each statement, at least half of all participants stated that they experience it at least sometimes.
Across the board, automotive participants experience the negative impact of test flakiness more often than global ones, an effect that we already observed in \prevalenceRQ.\@
The only exception is the option \textit{losing trust in passing tests to indicate the absence of regression} (5th row):
While global participants claim to suffer very similar amounts of trust loss towards both failing and passing test cases, automotive participants retain confidence in passing tests at a far higher rate while mistrusting failures much more frequently. This might be caused by global developers using more manual testing, where instinct plays a large role in finding hidden bugs.
Both participant groups agree that developer time is wasted more frequently than computational resources.%

When submitting code changes, automotive participants are specifically affected, with \percentAutomotiveConsequenceMergingRRsVeryoftenAlways stating that flaky tests hinder them in merging pull requests very often or always.
A likely explanation for this observation can be found in \cref{tab:demographic_results}, where we notice that almost all automotive participants use continuous integration, while only \SI{33}{\percent} of global developers do so.
Additionally, more than \SI{60}{\percent} of all participants see releases getting delayed at least sometimes due to test flakiness.

We also received 31 free text answers listing other consequences of test flakiness.
Many comments fall into the categories \textit{wasting developer time} (8x) and \textit{losing trust in testing} (7x), which confirms the selection of our pre-defined statements.
Both automotive and global participants also repeatedly mention frustration (5x) as a consequence of test flakiness.
Further responses state that flakiness %
results in production failures (4x), %
creates pressure by blocking or requiring last time changes (2x), and %
causes uncertainty if test failures actually indicate regressions (2x).
Interestingly, deactivating tests and writing rerun bots, which we consider mitigation strategies, were mentioned as negative consequences of flaky tests, indicating that the developers are dissatisfied with the way flaky tests are currently being handled.

To verify the existence of an \textit{asymmetric perception}~\cite{Machalica2020} of developers towards test executions, we looked at two aspects:
First, we checked if the failure rate of a flaky test has influence on how severely developers perceive it.
To measure this effect, we calculated the correlation between the two ordinal variables prevalence (question 7) and severity (question~8), using Kendall's tau.
We found a strong, positive, and significant connection between those variables %
($\tau = 0.39$, $p < 0.001$)%
, which confirms that flaky tests with high failure rates are also perceived as more problematic.
Second, we checked if developers lose trust in failing tests at the same rate as they do in passing tests.
To measure this effect, we verified if the responses of the two statements \textit{Rerun passing tests suspecting hidden bugs} and \textit{Rerun failing tests without analyzing the failure} originate from the same distribution.
Since a Shapiro-Wilk Test showed that their values are not normally distributed, we used a Wilcoxon signed-rank test, which resulted in a $p$-value of less than $0.001$, meaning that the two statements do not originate from the same distribution.
Together with the higher mean value of \textit{Rerun failures without analyzing} (\cref{tab:consequencesTable}), this shows that developers lose trust in the signaling power of test failures more than in the signaling power of passing tests.
However, there is a large difference between the two groups:
While this effect is extremely strong for automotive participants, %
it is not significant when looking only at global participants ($p = 0.13$).
This difference might be caused by the fact that automotive participants experience flaky failures more often, which is reported to make developers more likely to ignore genuine test failures~\cite{Parry2022}, or---as we suspect---rerun them.
Furthermore, only global participants mention production failures as a consequence of test flakiness, an experience that might lead them to also rerun passing tests.

Overall, we can confirm the existence of an asymmetric perception, showing that many developers look at test executions differently depending on the test's outcome.
A possible explanation for this phenomenon could be that most flakiness originates from the test code ($\sim$ \SI{90}{\percent} according to~\cite{Eck2019}) or the test environment, so most flaky tests are indeed false alarms and not hidden bugs.
However, while being rarer, flakiness caused by hidden bugs is arguably more devastating.
Considering flaky tests solemnly as false alarms is therefore a dangerous action, which cannot be recommended.
The asymmetric perception can therefore be seen as an indicator of the deceptive power flaky tests possess in masking actual bugs~\cite{Ziftci2020}.

\summary{\consequencesRQ: Negative Effects of Flaky Tests}{
    Flakiness destroys trust in testing and causes frustration.
    The degree of this also depends on the test outcome.
    Flaky tests inhibit project development by wasting developer time more than computational resources and blocking pull requests.
}

\subsection{\mitigationsRQ: Mitigation Strategies}
\label{sec:results_rq4}

\begin{table}[]
    \centering
    \caption{
        Mean rating per mitigation strategy; never (0) - always (4).
    }
    \label{tab:mitigations}
\begin{tabular}{lccc}
\toprule
{Mitigation strategies}        & {Global}                                           & {Auto}                                             & {Both} \\
\midrule
Rerun                          & \cellcolor[HTML]{292929} \color[HTML]{f1f1f1} 2.70 & \cellcolor[HTML]{000000} \color[HTML]{f1f1f1} 3.05 & \cellcolor[HTML]{1d1d1d} \color[HTML]{f1f1f1} 2.80 \\
Rerun in different environment & \cellcolor[HTML]{737373} \color[HTML]{f1f1f1} 2.10 & \cellcolor[HTML]{c6c6c6} \color[HTML]{000000} 1.37 & \cellcolor[HTML]{8a8a8a} \color[HTML]{f1f1f1} 1.89 \\
Auto report                    & \cellcolor[HTML]{a7a7a7} \color[HTML]{f1f1f1} 1.66 & \cellcolor[HTML]{dcdcdc} \color[HTML]{000000} 1.12 & \cellcolor[HTML]{b9b9b9} \color[HTML]{000000} 1.51 \\
Flag                           & \cellcolor[HTML]{b6b6b6} \color[HTML]{000000} 1.53 & \cellcolor[HTML]{bfbfbf} \color[HTML]{000000} 1.45 & \cellcolor[HTML]{b9b9b9} \color[HTML]{000000} 1.51 \\
Shuffle test order             & \cellcolor[HTML]{afafaf} \color[HTML]{000000} 1.59 & \cellcolor[HTML]{e5e5e5} \color[HTML]{000000} 1.00 & \cellcolor[HTML]{c2c2c2} \color[HTML]{000000} 1.43 \\
Incentives \& Penalties        & \cellcolor[HTML]{c5c5c5} \color[HTML]{000000} 1.39 & \cellcolor[HTML]{c4c4c4} \color[HTML]{000000} 1.40 & \cellcolor[HTML]{c5c5c5} \color[HTML]{000000} 1.40 \\
Disable                        & \cellcolor[HTML]{d8d8d8} \color[HTML]{000000} 1.18 & \cellcolor[HTML]{afafaf} \color[HTML]{000000} 1.59 & \cellcolor[HTML]{cdcdcd} \color[HTML]{000000} 1.29 \\
Quantify                       & \cellcolor[HTML]{c9c9c9} \color[HTML]{000000} 1.34 & \cellcolor[HTML]{dfdfdf} \color[HTML]{000000} 1.09 & \cellcolor[HTML]{cfcfcf} \color[HTML]{000000} 1.27 \\
Visualize                      & \cellcolor[HTML]{c9c9c9} \color[HTML]{000000} 1.34 & \cellcolor[HTML]{e1e1e1} \color[HTML]{000000} 1.04 & \cellcolor[HTML]{d1d1d1} \color[HTML]{000000} 1.26 \\
Auto detect                    & \cellcolor[HTML]{d4d4d4} \color[HTML]{000000} 1.21 & \cellcolor[HTML]{e3e3e3} \color[HTML]{000000} 1.02 & \cellcolor[HTML]{d9d9d9} \color[HTML]{000000} 1.16 \\
Auto debug                     & \cellcolor[HTML]{c9c9c9} \color[HTML]{000000} 1.34 & \cellcolor[HTML]{ffffff} \color[HTML]{000000} 0.53 & \cellcolor[HTML]{dcdcdc} \color[HTML]{000000} 1.11 \\
Auto disable                   & \cellcolor[HTML]{e2e2e2} \color[HTML]{000000} 1.04 & \cellcolor[HTML]{e4e4e4} \color[HTML]{000000} 1.01 & \cellcolor[HTML]{e2e2e2} \color[HTML]{000000} 1.03 \\
\bottomrule
\end{tabular}

\end{table}

\cref{tab:mitigations} shows the responses to the second last question of our survey, touching the mitigation strategies developers currently apply to deal with flakiness.
Rerunning tests is by far the most common reaction, while automated techniques rank lowest.
This might be the case because many sophisticated mitigation tools are implemented as plugins to Java build tools~\cite{Bell2018, Lam2019, Alshammari2021}, however, many industrial projects---especially in the automotive industry---do not use Java, preventing developers from applying these techniques without re-implementing them.
Furthermore, we received 20 free text answers and 6 responses to the last question (wishes), which also report currently applied strategies.
The majority of these describe rewriting the test case to remove its flakiness (11x), manually investigating the root cause (7x), and rerunning tests (5x).
Two developers see communicating with colleagues as a good strategy, namely talking to the test owner and raising awareness for memory errors.
One participant mentions that using Sonarlint~\cite{sonarlint}, an IDE extension that identifies and helps to fix quality and security issues, is useful in preventing more obvious cases of flakiness.

\summary{\mitigationsRQ: Currently Applied Mitigation Strategies}{
    Rerunning and rewriting test cases are by far the most dominant approaches to deal with flaky tests, while automated techniques are only rarely used.
}

\subsection{\wishesRQ: Wishes}%
\label{sec:results_rq5_wishes}

We received 187 suggestions and wishes for tools and information that developers would like to have in order to better deal with flaky tests. %
While automotive participants answered this question less often than their global counterparts
(23 / \automotiveNumNotFullyIncompleteSubmissions vs. 164 / \globalNumNotFullyIncompleteSubmissions),
their answers are more detailed (median 156 characters vs. 67 for global participants).
After coding the submissions, we discarded 30 responses from global participants because they were not interpretable (14x) or very generic like \enquote*{Don't know} (7x), \enquote*{Nothing} (5x), or \enquote*{Anything} (3x).
Furthermore, we found four answers from global participants that do not mention any wishes but instead describe root causes or mitigation strategies. We ignored these here but referred them to the respective RQ.\@
\cref{tab:wishes} summarizes the 153 meaningful, on-topic answers.

\begin{table}[]
    \centering
    \caption{Wishes for tools or information to better handle test flakiness expressed by our participants.}
    \label{tab:wishes}
    \begin{tabularx}{0.24\textwidth}{l@{\hskip 0.2cm}>{\hangindent=0.2cm}Xr}
    \toprule
        \multicolumn{2}{l}{Wish} & Count \\
    \midrule
        \multicolumn{2}{l}{\underline{Visualization}} & \underline{32} \\
        & test result history & 14 \\
        & which tests are flaky & 4 \\ %
        & program behavior & 3 \\  %
        \multicolumn{2}{l}{\underline{Auto Detection}} & \underline{31} \\
        & via static analysis & 12 \\
        \multicolumn{2}{l}{\underline{\smash{Auto Debug}}} & \underline{28} \\
        & find root cause & 6 \\
        & find location & 5 \\
        & find failure cause & 4 \\
        \multicolumn{2}{l}{\underline{Education}} & \underline{25} \\
        & guides, examples, best practices & 16 \\
        & help from colleagues & 3 \\
        & training & 2 \\
    \bottomrule
    \end{tabularx}
    \begin{tabularx}{0.243\textwidth}{l@{\hskip 0.2cm}>{\hangindent=0.2cm}Xr}
    \toprule
        \multicolumn{2}{l}{Wish} & Count \\
    \midrule
        \multicolumn{2}{l}{\underline{Rerun}} & \underline{17} \\
        & shuffle execution order & 4 \\
        & in different environment & 3 \\
        \multicolumn{2}{l}{\underline{\smash{Manual debugging tools}}} & \underline{16} \\
        \multicolumn{2}{l}{\underline{\smash{Stable infrastructure}}} & \underline{16} \\
        & machine readable infrastructure status & 5 \\
        \multicolumn{2}{l}{\underline{\smash{Logging}}} & \underline{13} \\
        \multicolumn{2}{l}{\underline{\smash{Management}}} & \underline{11} \\
        & priority + resources & 4 \\
        & good team & 4 \\
        & incentives \& penalties & 2 \\
        & issue bug reports & 2 \\
        \multicolumn{2}{l}{\underline{\smash{Reproducibility}}} & \underline{9} \\
        & failure replay & 3 \\
    \bottomrule
    \end{tabularx}
\end{table}

We find a strong desire for better visualization of flakiness, coming from both global participants (21x) as well as half of all the responses of automotive participants (11x)!
They specifically ask for tools to display the test result history, stating that \enquote*{We run tests so often, I often miss the bigger picture} (P~187), wishing for \enquote*{A tool to visualize how tests fail and succeed over time} (P~12). Such a tool should also include meta information like \enquote*{on which device / environment [the test was executed]} (P~292).

The second most commonly stated wish asks for tools that can automatically detect flaky tests.
Some respondents also go into details on how this should be implemented, stating that they would prefer static analysis-driven approaches, preferably an IDE plugin that can identify potentially flaky tests before executing them, similar to Findbugs~\cite{Hovemeyer2004}.
Many participants also want tools that can automatically debug flaky tests to determine the flakiness's root cause or location in the code.%
Various responses ask for more educational opportunities, specifically best practices and guides on how to avoid or fix flakiness, examples for flaky tests, and more support from forums like Stack Overflow.
Two participants would like to receive classes on test flakiness or suggest integrating the topic into programming courses.
Others mention that \enquote*{no tool can replace experience} (P~65) and wish for more support from (senior) colleagues.
One participant stated that they found our survey itself educational and plan to follow up on the topic.

The need for proper project management is raised by 11 participants, who ask for more (human) resources, skilled and loyal team members, and reporting of flaky tests with management overseeing the fixing progress.
One participant brought up the broken window theory, emphasizing the importance of preventing minor damage and conducting small repairs as fast as possible to avoid the establishment of a don't-care attitude.

Nine participants
 also mention the need for ways to reproduce a test's behavior, potentially through recording and replaying inputs.
Only two participants wish to know about the flakiness-introducing commit, which goes along Eck et al.'s~\cite{Eck2019} results seeing this information as less important.

\summary{\wishesRQ: Wishes for Information and Tools}{
    \begin{enumerate}
        \item Dashboards to visualize test outcomes over time
        \item IDE plugins to detect potentially flaky tests statically
        \item Automated debugging and root causing
        \item Education like examples, guides and best practices
    \end{enumerate}
}

\section{related work}%
\label{sec:related_work}

The closest related work to ours is a survey by Eck et al.~\cite{Eck2019} carried out in 2019 on 121 developers, in which
they found flakiness to be a frequent and severe issue, which our results corroborate.
Unlike them, however, we found connections between the domain of a project and the diffusion of flakiness in it.
Like ours, the participants of Eck et al.\ describe the main consequences of flakiness to be the waste of developer time and the decreased reliability and trustworthiness of a test suite, however, we saw no impact on resource allocation.
Similar to us, Eck et al.\ found a need for methods to prevent the introduction of flaky tests and for defining design patterns that support the creation of deterministic tests as well as flakiness-related anti-patterns.
On top of that, we saw a strong longing for better visualization techniques, such as dashboards showing the outcome of a test over time.
The usefulness of such overviews has been pointed out by grey literature~\cite{Oezal2021a, Champier2019, Liviu2019, Palmer2019}, however, it has not received much attention from researchers so far to the best of our knowledge.

Independently of our survey and at the same time, two other studies queried developers about test flakiness:
Ahmad et al.~\cite{Ahmad2021} interviewed 14 employees from five companies, two of which operate in the automotive business.
Like us, they found flakiness to be overproportionally common in the automotive sector, which might be caused by a relationship between flakiness and the context (i.e., domain) in which it has been investigated.
Furthermore, they also saw a strong need to prevent flakiness through guidelines and to predict it.
Parry et al.~\cite{Parry2022} surveyed 170 developers from various organizations about their experience of flaky tests.
Our results corroborate their findings that flakiness has a strong impact on CI, its root causes are often unknown, and that rerunning tests, as well as emotive responses like frustration, are common reactions.

Concerning the causes of flakiness, our survey confirms multiple previous studies that found concurrency to be the major cause of flakiness~\cite{Luo2014, Thorve2018, Eck2019, Lam2020}, the majority of order-dependent tests to be victims~\cite{Shi2019, Gruber2021}, and remote networking issues to be more common than local ones~\cite{Luo2014}.

\section{Conclusions}%
\label{sec:conclusions}

Seeking answers to the question of what researchers need to know about the developers' needs, desires, and experiences regarding test flakiness to build better defense mechanisms, we conducted an empirical study on 335 professional software developers and testers.
About one third of our participants originate from the automotive sector, allowing us to investigate the influence of domain-specific properties.

Flakiness is generally perceived as a prevalent and severe issue, especially within those using automated testing.
Mobile apps suffer from frequent flaky failures due to flaky emulators, whereas automotive platforms used for safety-relevant applications experience flaky failures less often but rate them as highly severe.
Concurrency-related issues remain the major causes of flakiness.
Furthermore, we found uninitialized variables, differences between compilers, and undocumented API changes as novel reasons for tests to become flaky.
Developers currently mostly rely on rerunning and rewriting test cases, however, their wishes towards researchers and tool developers are to provide more sophisticated techniques such as IDE plugins to detect potentially flaky tests or dashboards to visualize test outcomes over time.
Furthermore, they would like to receive more information and training on the topic.

To promote future research aiming to implement these needs or to replicate our work, we provide all data freely:\newline\figshareURL

\vfill
\noindent\textbf{Acknowledgements:}
We thank Philipp Straubinger for his support in coding free text answers.

\balance

\end{document}